\documentclass[structabstract]{aa}  

\usepackage{graphicx}
\pdfoutput=1
\usepackage[utf8]{inputenc}
\usepackage{txfonts}
\usepackage{natbib}
\bibpunct{(}{)}{;}{a}{}{,} % to follow the A&A style

\def\aap{A\&A\ }

\def\apjl{ApJL\ }

\def\apjs{ApJS\ }

\newcommand{\excs}{\extracolsep{\fill}}
\begin{document}

   \title{The young, tight and low mass binary TWA22AB: a new calibrator for evolutionary models ?\thanks{Based on service-mode observations (072.C-0644, 073.C-0469, 075.C-0521, 076.C-0554, 078.C-0510, 080.C-0581) collected at the European Organisation for Astronomical Research in the Southern Hemisphere, Chile.}}

   \subtitle{Orbit, spectral types and temperatures determination}

   \author{M. Bonnefoy\inst{1}
          , G. Chauvin\inst{1}
		  , C. Dumas\inst{2}
          , A-M. Lagrange\inst{1}
		  , H. Beust\inst{1}
		  , M. Desort\inst{1}
		  , R. Texeira	\inst{3}
		  , C. Ducourant\inst{4}
 		  , J-L. Beuzit\inst{1}
			\and
		   I. Song\inst{5}
          }

   \institute{Laboratoire d'Astrophysique de Grenoble,
              BP 53, F-38041 GRENOBLE C\'edex 9, France.\\
              \email{mickael.bonnefoy@obs.ujf-grenoble.fr}
         \and
             ESO, Alonso de Cordova 3107, Vitacura, Casilla 19001, Santiago 19, Chile.
			\and
Instituto de Astronomia, Geof\'isica e Ci\^encias Atmosf\'ericas,
       Universidade de S\~ao Paulo,
       Rua do Mat\~ao, 1226 - Cidade Universit\'aria,
       05508-900 S\~ao Paulo - SP,
       Brazil.
			\and
			 Observatoire Aquitain des Sciences de l’Univers, CNRS-UMR 5804, BP 89, 33270 Floirac, France.
			\and
			Department of Physics and Astronomy, University of Georgia, Athens, GA 30602, USA.
             }

%             \email{cdumas@eso.org}
%             \email{teixeira@astro.iag.usp.br}
%             \email{Christine.Ducourant@obs.u-bordeaux1.fr}
%             \email{song@uga.edu}

%\offprints{mickael.bonnefoy@obs.ujf-grenoble.fr}

   \date{Received September 08, 2008; accepted May 17, 2009}

\authorrunning{Bonnefoy et al.}
\titlerunning{The young, tight and low mass binary TWA22AB: a new calibrator for evolutionary models ?}

% \abstract{}{}{}{}{} 
% 5 {} token are mandatory
 
  \abstract
   {Tight binaries discovered in young, nearby associations, with
   known distances, are ideal targets to provide dynamical mass
   measurements to test the physics of evolutionary models at young
   ages and very low masses.}
   {We report for the first time the binarity of TWA22. We aim at monitoring the orbit of this young and tight system to
   determine its total dynamical mass using an accurate
   distance determination. We also intend to characterize the physical
   properties (luminosity, effective temperature and surface gravity)
   of each component based on near-infrared photometric and
   spectroscopic observations. %The final issue is to test if TWA22\,AB can be used as a new calibrator for evolutionary model predictions  at young ages and at very low stellar masses.
%AML m'a dit d'enlever cette phrase.
% and thus to test evolutionary models predictions for young objects, assuming that the system is a member of the 8 years old TW Hydrae association.
}
   {We use the adaptive optics assisted imager NACO to resolve the components, to monitor the
   complete orbit and to obtain the relative near infrared photometry
   of TWA22 AB. The adaptive optics assisted integral field
   spectrometer SINFONI was also used to obtain medium resolution ($R_{\lambda}=1500-2000$) 
   spectra in JHK bands. Comparison with empirical and
   synthetic librairies were necessary to derive the spectral type, the
   effective temperature and the surface gravity for each component of
   the system.}
   {Based on an accurate trigonometric distance (17.53
   $\pm$ 0.21 pc) determination, we infer a total dynamical mass
   of 220 $\pm$ 21 M$_{Jup}$ for the system. From the complete set of spectra, we find an effective temperature $T_{eff}=2900^{+200}_{-200}$ K for TWA22 A and $T_{eff}=2900^{+200}_{-100} $ K for TWA22 B and surface gravities between 4.0 and 5.5 dex. From our photometry and a M6 $\pm$ 1
   spectral type for both components,
we find luminosities of log(L/L$_{\odot}$)=-2.11$ \pm$ 0.13 dex and
   log(L/L$_{\odot}$)=-2.30$\pm$ 0.16 dex for TWA22 A and B respectively. By
comparing these parameters with evolutionary models, we question the age and the multiplicity  of this system. We also discuss a possible underestimation of the mass predicted by evolutionary models for young stars close to the substellar boundary.}
   {}

\maketitle
%
%________________________________________________________________

\section{Introduction}

\begin{table*}[t]
\begin{minipage}[t]{\linewidth}
\caption{Observing log.}             % title of Table
\label{tab:table1}
\centering
\renewcommand{\footnoterule}{}  % to avoid a line before footnotes
\begin{tabular*}{\textwidth}{@{\excs}lllllllllll}     % 7 columns
\hline\hline
 UT Date    & Name      &      Instrument    &   Mode            &Filter    &   Camera     &     Airmass   &   Seeing       &   EC  \footnote{Corresponds to the mean strehl ratio for the spectroscopic observations.}    & Exp.    &   Note     \\ 
            &           &                    &                  & (Grism) &              &               &                &             &       Time         &             \\ 
            &           &                    &                  &  &              &               &        (arcsec)        &        (\%)        &      (s)          &             \\ 
\hline
2004/03/05  & TWA22AB    & NACO          & imaging   &   \textit{NB\_2.17}        & S27          &      1.16    &      1.23  	         &     		16.5	         &        20        &            \\  
2004/03/05  & TWA22AB    & NACO	& imaging		  &   \textit{NB\_1.24}		  & S13			&			1.15		&	    1.40      	         &     	27.7		         &        120        &            \\  
2004/04/27  & TWA22AB    & NACO	& imaging		  &   \textit{NB\_1.75}		  & S27			&			1.16		&	    0.84              &     		48.2         &        6        &            \\  
2005/05/06  & TWA22AB    & NACO	& imaging		  &   \textit{H}		  		   	  & S13			&			1.15		&	    0.72              &     		12.5	         &       5         &            \\  
2006/01/08  & TWA22AB    & NACO& imaging			  &   \textit{J}					  & S13			&			1.16		&	    0.84              &     	 17.5	         &      50          &            \\  
2006/01/08  & TWA22AB    & NACO	& imaging		  &   \textit{H}					  & S13			&			1.17		&	    0.65              &     		17.4	         &       50         &            \\  
2006/02/26  & TWA22AB    & NACO	& imaging		  &   \textit{H}					  & S13			&		1.15			&	    1.10              &     			32.1         &        150        &            \\  
2006/02/26  & TWA22AB    & NACO	& imaging		  &   \textit{K}$_{s}$		  & S27			&			1.15		&	    1.50               &     	   10.2	        &       50         &            \\
2007/03/06  & TWA22AB    & NACO	& imaging		  &   \textit{H}					  & S13			&		1.15			&	    0.60               &     		45.7	         &      103.5          &            \\ 
2007/12/04  & TWA22AB    & NACO	& imaging		  &   \textit{H}					  & S13			&		1.28		&	    1.00               &     		34.4	         &        150        &            \\     
2007/12/26  & TWA22AB    & NACO	& imaging		  &   \textit{H}					  & S13			&		1.16			&	    0.90               &     		39.9	         &      150          &            \\     
2007/02/12  & TWA22AB    & SINFONI& spectroscopy         &   \textit{J} (2000)           & 25           &     1.18      &     1.56           &       12.1         &         1080       &            \\  
2007/02/13  & TWA22AB    & SINFONI & spectroscopy        &  \textit{J} (2000)           & 25           &     1.16      &     0.98           &       18.5         &       1080         &            \\
2007/02/13  & HIP049201  & SINFONI & spectroscopy        &   \textit{J} (2000)           & 25           &     1.13      &     0.86           &       27.0         &       20         &       Telluric Standard    \\
2007/02/13  & HIP038858  & SINFONI & spectroscopy        &   \textit{J} (2000)           & 25           &     1.14      &     0.78           &       28.5         &        20        &      Telluric  Standard     \\
2007/02/12  & HIP035208  & SINFONI & spectroscopy        &   \textit{J} (2000)           & 25           &     1.17      &     1.17           &       15.2         &       60         &      Telluric Standard      \\
%2007/02/11  & LHS 337      & SINFONI         &   J (2000)            & 25 			&	   1.07	 	&      0.81 		  & 		26.49	 	 &        1080        &        PSF reference    \\
2007/02/11  & TWA22AB    & SINFONI  & spectroscopy       &   \textit{H+K} (1500)      & 25           &     1.16		&  	0.89           &       29.5		 &			960		&		   	  \\
2007/02/09  & TWA22AB    & SINFONI& spectroscopy		 &   \textit{H+K} (1500)		  & 25			& 		1.52     &      1.00             &       17.2        &		960			&		   	  \\
2007/02/11  & HIP052202  & SINFONI& spectroscopy		 &   \textit{H+K} (1500)		  & 25			& 		1.19     &     0.93            &       24.7        &		20			&		Telluric Standard    	  \\
2007/02/09  & HIP052202  & SINFONI& spectroscopy		 &   \textit{H+K} (1500)		  & 25			& 		1.55     &     1.22            &       14.0        &		40			&		Telluric Standard   	  \\
%2007/02/10  & LHS 337      & SINFONI		 &   H+K (1500)		  & 25			& 		1.07     &     0.62            &       32.69        &		960			&		   	  PSF reference \\

\hline
\end{tabular*} 
\end{minipage}
\end{table*}

Mass and age are fundamental parameters of stars and brown
dwarfs that determine their luminosity, effective temperature,
atmospheric composition and surface gravity commonly derived through
photometric and spectroscopic observations. Evolutionary models are
currently widely used in the community to infer masses of stars and
brown dwarfs, but they rely on equations of states and atmospheric
models non calibrated at young ages and at very low
masses. However, direct mass measurements
can be obtained by the mean of different observing
techniques, combining light curve studies to radial velocity
on eclipsing binaries, astrometric follow-up with double lined
spectroscopy of tight binaries or measuring the Keplerian
motion of circumstellar disks. In recent years, direct
mass measurements for 23 pre-main sequence stars with masses ranging
from 0.5 to 2 $M_{\odot}$ showed
discrepancies with predictions by up to a factor of 2 in mass and 10
in ages \citep{2007prpl.conf..411M}.  Such measurements are more rare for lower masses ($\leqslant
0.5 M_{\odot}$) systems. \cite{2004ApJ...604..741H} showed that the
models tend to understimate the mass of the companion UZ Tau Eb
(M=0.294$\pm$0.027\,$M_{\odot}$, age $\backsim$5~Myr, \cite{2002ApJ...579L..99P}). \cite{2005Natur.433..286C} derived similar
conclusions but the age and the luminosity of the companion
AB Dor C (M=0.090$\pm$0.005
$M_{\odot}$, age$\backsim$75~Myr) is still under
debate \citep{2008A&A...482..939B}. And recently, the surprising
discovery of the unpredicted temperature reversal
\citep{2007ApJ...664.1154S} between 2M035 A (M=0.0541$\pm$0.0046
$M_{\odot}$, age $\backsim$1~Myr) and its companion
(M=0.0340$\pm$0.0027 $M_{\odot}$, age $\backsim$1~Myr) proves the
necessity to find more calibrators. The challenge is to determine
unambiguously their physical properties (mass, \textit{L}, $T_{eff}$, g and age) and
to explore as much as possible the parameter space covered by
evolutionary models. The influence of other parameters such as
metallicity needs also investigation \citep{2005ApJ...635..442B, 2007AJ....134.1330B}.

The TW Hydrae association (TWA) is the first co-moving group of young
($\le 100$~Myr), nearby ($\le 100$~pc) stars, that was
identified near the Sun \citep{1997Sci...277...67K}. Ideal
observational niche for the study of stellar and planetary formation,
TWA was actually the tip of an iceberg composed of hundreds of young
stars, spread in different groups, that were identified during the
last decade \citep{2004ARAA..42..685Z, 2008arXiv0808.3362T}. TWA counts now 27
members covering a mass regime from intermediate-mass stars to
planetary mass objects \citep{2005AA...438L..25C}. Its 8.3 $\pm$ 0.8 Myr dynamical age was found by a convergence method \citep{2006AJ....131.2609D}. Independently, \cite{2006AA...459..511B} estimated an age of $10^{+10}_{-7}$ Myr from the photometry, the activity and the lithium depletion. \cite{2007ApJ...662.1254S} show that the association is $9^{+8}_{-2}$ Myr old by comparing rotational velocities with published rotation periods for a subset of stars. Finally, \cite{2008ApJ...689.1127M} estimated an age of 12 $\pm$ 8 Myr from the study of lithium depletion in five nearby young associations (hereafter M08).
%\cite{2003ApJ...599..342S} used visual spectroscopy (strong Li
%absorption, $H_{\alpha}$ emission) and proper motions to identify
%TWA22 as an M5 member of TWA. The proximity of TWA22 (17.53 $\pm$ 0.21
%pc, see Ducourant et al. 2008) made it ideal to search for short
%revolution period and low inclination companions with the use of
%recent developed high angular resolution techniques.}

\begin{figure}[t]
   \centering \includegraphics[width=6cm]{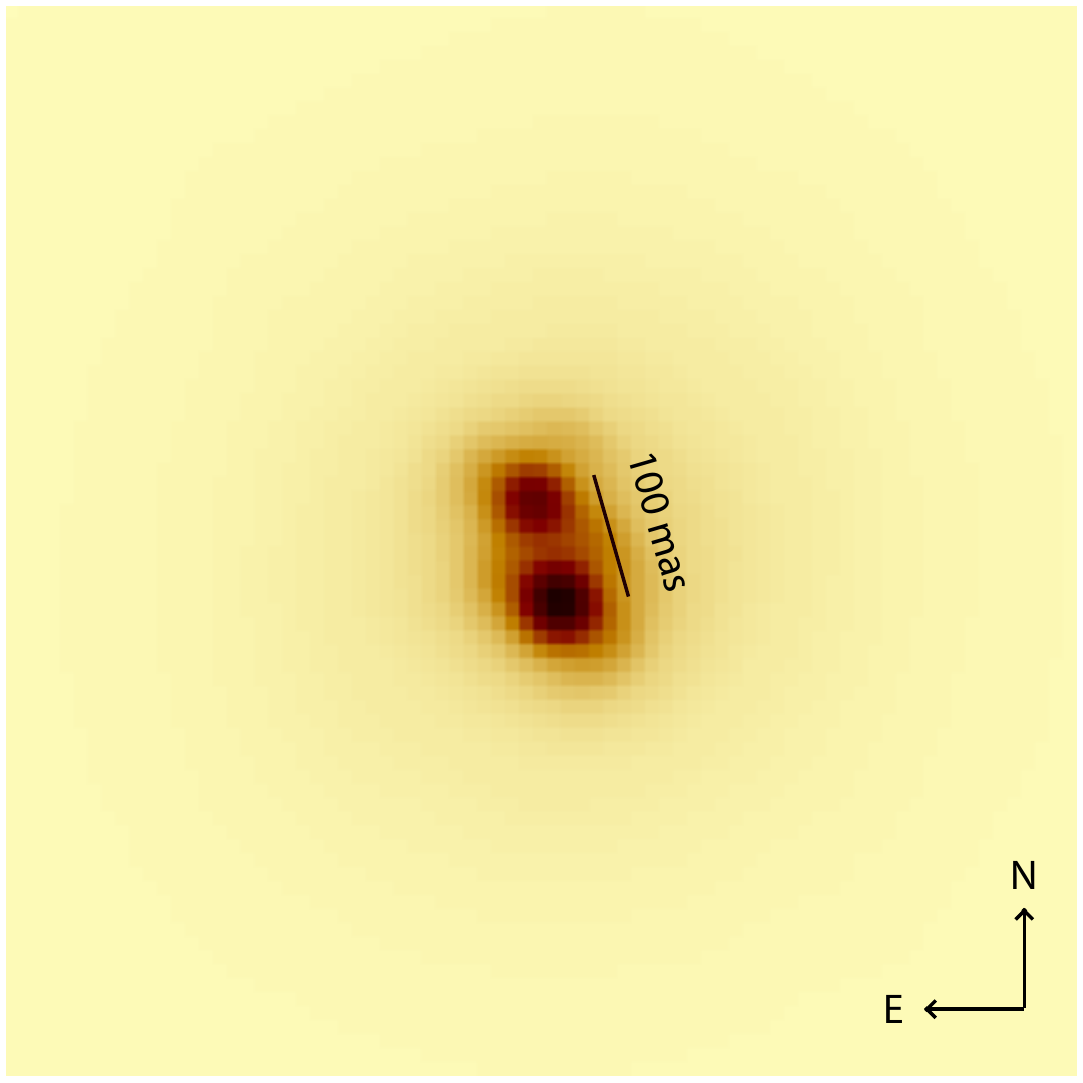} 
     \caption{VLT/NACO image of TWA22~AB obtained in H-band with the
     S13 camera  on December 26th, 2007. North is up and East is left. The field of view is
     $1~\!'' \times1~\!''$.}
         \label{Image_Twa22AB}
\end{figure}

\cite{2003ApJ...599..342S}, hereafter S03, identified TWA22 as an M5 member of TWA. The strong Li $\lambda$6708 \AA~feature supported the extreme youth of this member. Later, \cite{2005ApJ...634.1385M} questioned the membership of TWA22 from a kinematic study of TW Hydrae members. Finally, \cite{2006ApJ...652..724S} discussed the  \cite{2005ApJ...634.1385M} results that appeared in disagreement with the very strong lithium line of the source. The proximity (17.5 $\pm$ 0.2 pc, see \cite{2009Teixeira}) and the reported youth of TWA22 by S03 made it consequently a potential target for the detection of companions at small orbital radii.

In our program for detection of companions in young associations using the adaptive optic (AO) assisted imager NACO, we resolved TWA22 as a tight ($\sim$100~mas)  binary. With a projected physical separation of $1.76\pm0.10$~AU (see Fig. \ref{Image_Twa22AB}),  this system offered a unique opportunity to measure its dynamical mass and to possibly test the evolutionary model predictions at young ages using combined photometric and spectroscopic observations.

We report here the discovery of the TWA22 binarity and the results of a dedicated 4 years observing program, using combined
imaging and 3D-spectroscopy with AO. The purpose
was to measure the dynamical mass of TWA22~AB and to characterize the
physical properties of the individual components. In §2, we describe our
AO observations with the VLT/NACO imager and with the VLT/SINFONI
integral field spectrograph. The associated data reduction and
spectral extraction techniques are detailed in §3. In §4, we present
our orbital solutions and our spectral analysis. In §5 we
compare and discuss the evolutionary model predictions associated to
our dynamical mass measurement with the physical properties (surface
gravity, temperature and luminosity) derived from our
photometric and spectroscopic observations. This leads us to discuss the
membership of TWA22\,AB to the TW Hydrae association, the multiplicty of the system and a
possible underestimation of the mass predicted by evolutionary models
for young stars close to the substellar boundary.

%__________________________________________________________________

\section{Observations}

\subsection{VLT/NACO Observations}

TWA22 AB was observed at the 8.2 m VLT UT4 Yepun with the Nasmyth
Adaptive Optic (AO) system NAOS \citep{2000SPIE.4007...72R} coupled to
the High-Resolution Near-IR Camera CONICA
\citep{1998SPIE.3354..606L}. NAOS and CONICA (NACO) resolved the system as a tight binary for the first time on March 5th, 2004. Follow-up observations were conducted during
four years from early 2004 to end 2007. To image TWA22\,AB, we used
the narrow band filters: \textit{NB\_1.24} ($\lambda_{0}$=1.237 $\mu$m,
FWHM=0.015 $\mu$m), \textit{NB\_1.75} ($\lambda_{0}$=1.748 $\mu$m, FWHM=0.026
$\mu$m), \textit{NB\_2.17} ($\lambda_{0}$=2.166 $\mu$m, FWHM=0.023 $\mu$m).
 The broad band filters
\textit{J} ($\lambda_{0}$=1.27 $\mu$m, FWHM=0.25 $\mu$m), \textit{H} ($\lambda_{0}$=1.66
$\mu$m, FWHM=0.33 $\mu$m) and $K_{s}$ ($\lambda_{0}$=2.18 $\mu$m,
FWHM=0.35 $\mu$m) were also used coupled to a neutral density
(attenuation factor of 80).  CONICA was used with the S13 and S27
cameras to Nyquist-sample the PSF depending on the selected
filter. The data were recorded under seeing ranging from $0.6~\!''$ to
$1.5~\!''$ (see Table \ref{tab:table1}). TWA22\,AB was bright enough in visible to be used by NAOS
for wave-front analysis. For each observation period, dithering around
the object in \textit{J}, \textit{H} and $K_{s}$ bands combined with nodding were
necessary to run a good sky estimation during the data reduction
process (see part 2.2). PSF references were observed at different
airmasses with identical setups. The $\Theta_1$ Ori C astrometric
field \citep{1994AJ....108.1382M} was observed at each epoch to
calibrate the detector platescale and orientation whenever
necessary. The results are reported in Table~\ref{tab:table2}.

% Texte de la section 3.1
%The position of stars in the $\theta$ Orioni astrometric field is
%compared to accurate positions obtained from HST images at 2.1 $\mu$m
%\citep{1994AJ....108.1382M}. Since the absolute orientation of CONICA
%is known with a $0.2^{\circ}$ accuracy, we performed only a relative
%astrometric calibration between different epochs. The relative
%position of Twa22B is given in table \ref{table:1}.

% Texte de la section 4.1
%For periods with only one reference PSF in the 13.27 mas/pixel
%platescale, \textit{Starfinder} extracted the positions of TWA22B in 1
%iteration indendantly. The difference between these results give an
%estimation of the error.  Deconvolution was not functionning well for
%27.15 mas/pixel platescale and with observations under bad seeing
%conditions.The influence of sub-images sizes used for deconvolution
%was probed and provided different positions depending of the extent of
%PSF wings. We concerved positions given by a minimum number of
%iterations.

\begin{table}[ht]
\caption{Mean plate scale and detector orientation for our different observing NACO runs.}
\label{tab:table2}
\centering
\begin{tabular*}{\columnwidth}{@{\excs}cccc}     % 7 columns
\hline\hline
 UT Date     & Camera       & Plate Scale    &    Orientation of true north  \\
&&& east of the vertical \\
             &              & (mas/pixel)          &    (o)                   \\
\hline
2004/03/05   & S27          & $27.01\pm0.05$     & $-0.18\pm0.19$    \\
2004/04/27   & S27          & $27.01\pm0.10$     & $0.08\pm0.20$    \\
2005/05/06   & S13          & $13.25\pm0.08$     & $-0.02\pm0.10$    \\
2006/01/08   & S13          & $13.26\pm0.08$     & $0.18\pm0.10$    \\
2006/02/26   & S13          & $13.25\pm0.06$     & $0.19\pm0.11$    \\
2007/03/06   & S13          & $13.28\pm0.10$     & $0.08\pm0.15$    \\
2007/12/04   & S13          & $13.29\pm0.12$     & $-0.21\pm0.17$    \\
2007/12/26   & S13          & $13.29\pm0.12$     & $-0.21\pm0.17$    \\
\hline
\end{tabular*}
\end{table}

\subsection{VLT/SINFONI observations}

The SINFONI instrument (Spectrograph for INtegral Field Observations in the Near Infrared),
located at the Cassegrain focus of the VLT UT4 Yepun, was used to
observe TWA22\,AB between February 9 and 13 2007. SINFONI includes an integral field spectrometer SPIFFI (SPectrograph for Infrared
Faint Field Imaging, see \citet{2003SPIE.4841.1548E}), operating in
the near-infrared (1.1 - 2.45 $\mu$m). SPIFFI is assisted with the 60
actuators Mutlti-Applications Curvature Adaptive Optic system MACAO
\citep{2003SPIE.4839..329B}. We used the small SPIFFI field of
view of 0.8" $\times$ 0.8" corresponding to a plate scale of 25 mas per pixel to
Nyquist-sample the SINFONI AO corrected PSF. The field of view is
optically sliced into 32 horizontal slitlets, that sample the horizontal spatial direction and which are rearranged to
form a pseudo-long slit.  Once dispersed by the grating on the 2048 $\times$ 2048 SPIFFI detector, each slitlet of 64 pixels width (spatial direction) corresponds to 64 spectra of 2048 pixels long (spectral direction). The 2048
independent spectra on the detector are reorganized during the reduction process in a
datacube which contains the spatial (X, Y) and the spectral (Z)
informations. The cube is resampled in the vertical dimension (Y) to
have the same number of pixels as in X.  

To cover the full spectral range between 1.1
to 2.45 $\mu$m, individual integrations times of 90 s were necessary to image the system in the \textit{J}
band  (1.1 - 1.4 $\mu$m,
R=2000) and 20 s in the \textit{H+K} band (1.45 - 2.45 $\mu$m, R=1500). For each band, dithering around the
object was used to increase the field of view and to supress residual bad pixels, leading to a total observing
time on target of $\sim$ 5 min. An additional frame was
acquired on the sky to improve our correction. The adaptive optic loop
was locked on TWA22\,AB itself. Standard stars HIP038858 (B3V),
HIP049201 (B2V), HIP035208 (B3V) and HIP052202 (B4V) were observed at
similar airmasses to remove the telluric lines (see
Table~\ref{tab:table1}).

\section{Data reduction and analysis}

\subsection{High Contrast Imaging}

For each observation periods, the ESO \textit{eclipse} reduction
software \citep{1997Msngr..87...19D} dedicated to AO image processing
was used on the complete set of raw images. Eclipse computes bad-pixel
detection and interpolation, flat field correction and averaging pairs
of shifted images with sub-pixel accuracy. The software run sky
estimation on object-dithered frames using median filtering through
the frame sequence.

 A deconvolution algorithm dedicated to stellar field blurred by the
adaptive optics corrected point spread functions
\citep{1998SPIE.3353..426V} was applied on TWA22 AB images to
accurately find the position and the photometry of the companion relative
to the primary.  The algorithm is based on the minimization in the
Fourier domain of a regularized least square objective function using
the Levenberg-Marquardt method. We used Nyquist-sampled unsaturated
images of standard stars obtained the same night as TWA22 observations
with identical setups under various atmospheric conditions. These
frames captured the variation of AO corrections. They were used as
input point spread functions (PSF) to estimate the deconvolution
process error. The IDL \textit{Starfinder}\footnote{\small IDL
procedures can be downloaded at http://www.bo.astro.it/$\sim$giangi/StarFinder/index.htm } PSF fitting
package \citep{2000SPIE.4007..879D} confirmed these results.

\begin{table}[b]
\caption{Relative positions and contrasts of TWA 22 A and B. Magnitude differences are given in the NACO photometric system.}             % title of Table
\centering
\label{tab:table3}      % is used to refer this table in the text                          % used for centering table
\renewcommand{\footnoterule}{}  % to avoid a line before footnotes
\begin{tabular*}{\columnwidth}{@{\excs}llllll}     % 7 columns
\hline \hline                  % inserts double horizontal lines
UT Date    & Filter     &  Camera   & $\Delta\alpha_{J2000}$ & $\Delta\delta_{J2000}$ & $\Delta m$  \\    % table heading 
           &            &           & (mas)                  & (mas)                  & (mag)  \\
\hline                      % inserts single horizontal line
2004/03/05 & NB2.17     & S27       & 99 $\pm$ 3             & -17 $\pm$ 3            & 0.33 $\pm$ 0.12	 \\
           & NB1.24     & S13       &                        &                        & 0.37 $\pm$ 0.10 \\
2004/04/27 & NB1.75     & S13       & 98 $\pm$ 6             & -36 $\pm$ 6            & 0.41 $\pm$ 0.10 \\
2005/05/06 & H-ND       & S13       & 15 $\pm$ 3             & -89 $\pm$ 3            & 0.66 $\pm$ 0.05 \\
2006/01/08 & J          & S13       &                        &                        & 0.40 $\pm$ 0.04 \\
           & H          & S13       & -68 $\pm$ 2            & -49 $\pm$ 2            & 0.54 $\pm$ 0.05 \\ 
2006/02/26 & H          & S13       &  -74 $\pm$ 3           & -30 $\pm$ 3            & 0.54 $\pm$ 0.05  \\
           & K$_{s}$    & S27       &                        &                        & 0.46 $\pm$ 0.18 \\
2007/03/06 & H          & S13       & -57 $\pm$ 4            & 80 $\pm$ 2             & 0.49 $\pm$ 0.03 \\
2007/12/04 & H          & S13       & 19 $\pm$ 3             & 98 $\pm$ 3             & 0.52 $\pm$ 0.10 \\
2007/12/26 & H          & S13       & 26 $\pm$ 3             & 97 $\pm$ 3             & 0.53 $\pm$ 0.10 \\
\hline                                   %inserts single line
\end{tabular*}
\footnotetext[1]{Seeing evaluated at 0.55 $\mu$m.}
\end{table}

\subsection{Integral Field Spectroscopy}

We used the SINFONI data reduction pipeline (1.7.1 version, see
\cite{2007astro.ph..1297M}) for raw data processing. The pipeline
carries out cube reconstruction from raw detector images. The flagging
of hot and non-linear pixels is executed in a similar way as in NACO
images. The distortion and wavelength scale are calibrated on the
entire detector using arc-lamp frames. Slitlet distances are
accurately measured with North-South scanning of the detector
illuminated with an optical fiber. In the case of standard stars
observation, object-sky frame pairs are subtracted, flat fielded and
corrected from bad pixels and distortions. Datacubes are finally
reconstructed from clean science images and are
merged in a master cube. Spectra of standard stars cleaned
from stellar lines are finally used to correct the TWA22 AB
spectra from telluric absorptions.

TWA22 A and B are centered and oriented horizontally in the \textit{J} and \textit{H+K}
master cubes with a field of view of
$1.1~\!''\times1.1~\!''$. Atmospheric refraction induces different
sources positions for different wavelengths within the instrument
field of view and increases with airmass. Combined with the small
SINFONI field of view, this produces differential flux losses that were
noticed in the bright standard stars datacubes. This
effect remains limited for TWA22.
The cubes of 11 February 2007 appear to have some spaxels contaminated
by flux oscillations of a few ADUs. These oscillations were not negligible and blurred CO
bands at 2.3 $\mu$m. They are present along
the dispersion axis in the raw detector images of both  HIP052202
and TWA22. Their amplitudes do not remain constant in time
but follow a 15.3 pixels period. We then filtered partially this contribution on each individual
image in the Fourier space using a pass-band function. The origin of
the problem is likely to be related to 50 Hz pick-up noise.

We used a modified version \citep{2001AJ....121.1163D} of the CLEAN
algorithm \citep{1974A&AS...15..417H,1978A&A....65.345M} to extract
separately the flux of TWA22 A and B in each monochromatic images
contained in the datacubes. The standard star are used as initial
PSF-references. Once scaled to match the TWA22 A maximum at the
primary position and for all wavelengths, the PSF is subtracted to the
TWA22 AB datacube. The sequence is repeated to model the secondary
contribution, cleaned from the primary wings, and to provide a new
PSF-reference. After a few iterations minimizing the final quadratic
residual datacube, the spectra of each individual component are
extracted.

The algorithm was first adapted to work on cube images. Unfortunately,
the difference of sampling between the X and Y directions limited the
sub-pixel shift accuracy. We therefore collapsed the cube along the
Y-axis in order to obtain the flux profile along the X direction.  We
chose to duplicate the primary flux profile for the PSF model. The
algorithm converged in a few iterations and produced extracted spectra
in\textit{ J} and \textit{H+K} with an extraction error less than
5\%. The extracted spectra were divided by standard star spectra
corrected from intrinsic features and multiplied by a black body
spectrum at the standard star temperature. The SINFONI pipeline
coefficients were used for wavelength calibration.

%__________________________________________________________________

\section{Results}

\subsection{Astrometry, Orbit and dynamical mass}

The relative positions of TWA 22 A and B (B with respect to A)
at all observation epochs are reported in
Table~\ref{tab:table3}. The data allows a determination of the
mutual orbit of the binary. We define a
cartesian referential frame $(O,X,Y,Z)$ where $X$ points towards the
north, $Y$ toward the east and $Z$ toward the Earth. The $(OXY)$ plane
corresponds thus to the plane of the sky. Then in a Keplerian
formalism, the $(x\equiv\Delta\delta,y\equiv\Delta\alpha)$
projected position of the binary onto the plane
of the sky reads
\begin{eqnarray}
\lefteqn{
x \; = \; \frac{a}{d}\,(\cos u-e)\left(\cos\omega\cos\Omega
-\sin\omega\cos i\sin\Omega\right)}&&\nonumber\\
 &&\mbox{}+\frac{a}{d}\,\sqrt{1-e^2}\sin u\left(-\sin\omega\cos\Omega
-\cos\omega\cos i\sin\Omega\right)\;,\\
\lefteqn{y \; = \; \frac{a}{d}\,(\cos u-e)\left(\cos\omega\sin\Omega
+\sin\omega\cos i\cos\Omega\right)}&&\nonumber\\
&& \mbox{}+\frac{a}{d}\,\sqrt{1-e^2}\sin u\left(-\sin\omega\sin\Omega
+\cos\omega\cos\Omega\cos i\right)\;,
\end{eqnarray}
where $a$ is the semi-major axis of the orbit (in AU), $d$ is the
distance of the binary (in pc), $e$ is the eccentricity, $i$ is the
inclination, $\Omega$ is the longitude of ascending node (counted from
north towards east), $\omega$ is the argument of periastron, and $u$
is the eccentric anomaly that describes the current location of the
binary along its orbit. $u$ is related to the time $t$ by the
classical Kepler's equation 
\begin{equation}
\frac{2\pi}{T}(t-t_\mathrm{p})=u-e\sin u\qquad,
\end{equation}
where $T$ is the orbital period and $t_\mathrm{p}$ is the time reference for
periastron passage. Once the distance $d$ is known (17.5\,pc,
\cite{2009Teixeira}), the fit of the observational data allows to
determine the 7 parameters $T$, $a$, $e$, $i$, $\Omega$, $\omega$ and
$t_\mathrm{p}$. Then Kepler's third law leads to the determination of
the total mass $M$.

The fit is performed via a Levenberg-Marquardt $\chi^2$ minimizing
algorithm. In practice, instead of $(e,i,\Omega,\omega)$, the
equations are solved for the classical variables
\begin{eqnarray}
k\;=\;e\,\cos(\Omega+\omega) && q\;=\;\sin\frac{i}{2}\,\cos\Omega\nonumber\\
h\;=\;e\,\sin(\Omega+\omega) && p\;=\;\sin\frac{i}{2}\,\sin\Omega\;.
\end{eqnarray}
which avoids singularities towards small eccentricities and inclinations.
The uncertainties on the fitted parameters are estimated from the
resulting covariance matrix at the end of the fit procedure. 

\begin{figure}[t]
   \centering
   \includegraphics[width=\columnwidth]{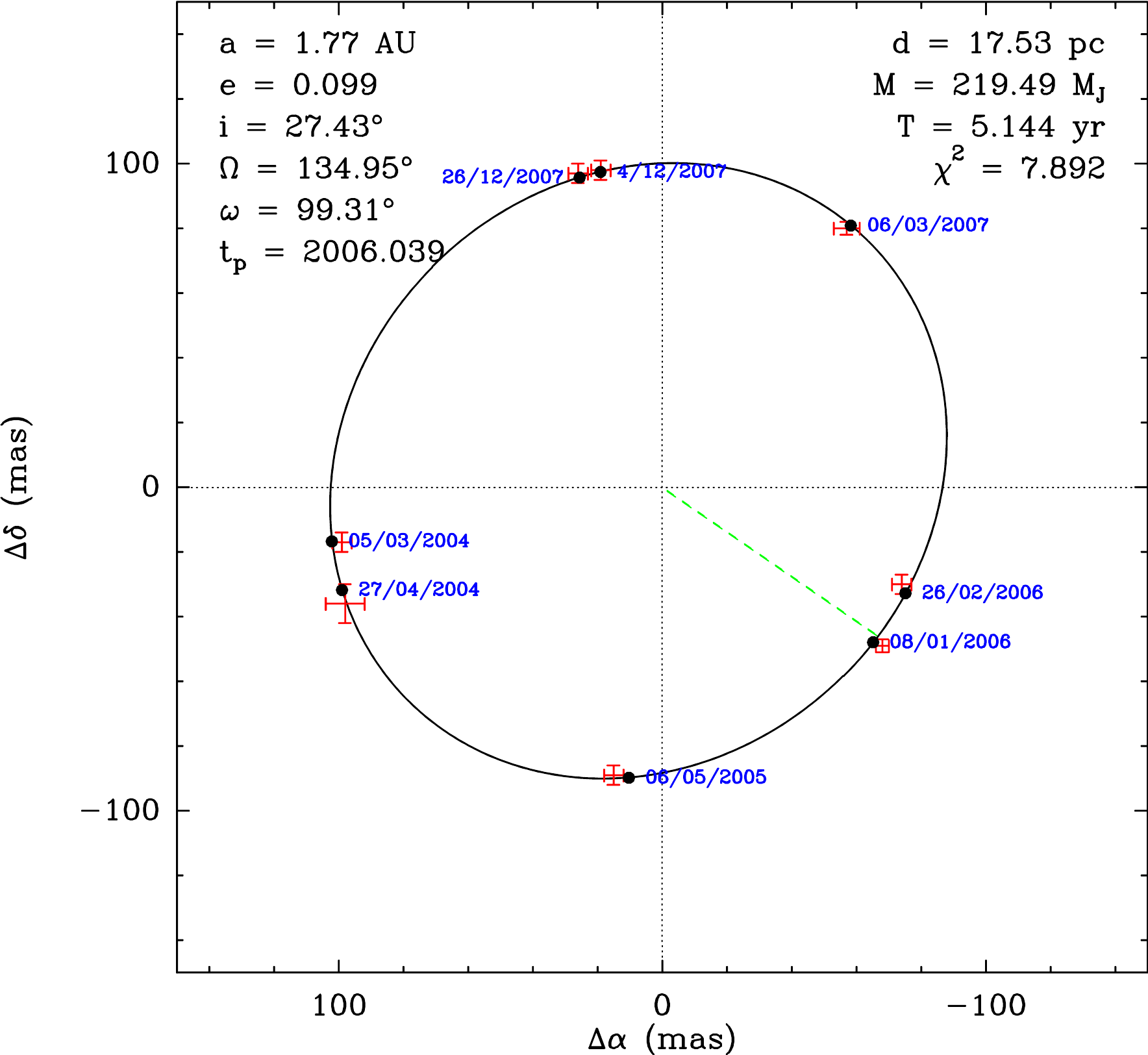}
      \caption{Orbital fit of the relative positions of TWA22\,AB
      observed from March 2004 to December 2007, as projected onto the
      plane of the sky. The crosses represent
      the observational data with their error bars, the solid line is
      the fitted projected orbit, and the dots correspond to the predicted
      positions of the model at the times of the observations. The
      dashed line sketches the projected direction of the periastron
      of the orbit. On Jan. 8, 2006, the binary was actually very
      close to periastron.}
         \label{orbit}
   \end{figure}

Levenberg-Marquardt is an interative gradient method for converging
towards a mininum of the $\chi^2$ function. Depending on the starting
guess point, many local minima can be found. In the present case,
\emph{all} the attemps we made (by letting the starting point vary)
converge towards the \emph{same}
solution that is listed in Table~\ref{Orbit_par} and viewed in
projection onto the plane of the sky in Fig.~\ref{orbit}. The
available astrometric data set appears to cover almost one complete
orbital period with a good sampling of the periastron passage. We are
thus confident in our fitted solution. The orbit appears then slighly
eccentric ($e\simeq0.1$) and viewed close to pole-on $i\simeq 27\degr$
from the Earth.

\begin{table}[h]
\caption{TWA22 B orbital parameters as determined from the fit of the
  astrometric data (see text for the definition of the parameters). The reduced $\chi^{2}$ of the fit is also reported.}             % title of Table
\label{Orbit_par}      % is used to refer this table in the text                          % used for centering table
\centering                          % used for centering table
\begin{tabular}{l@{\hspace{1truecm}}l}     % 7 columns
\hline     \hline
\noalign{\smallskip}
Reduced $\chi^{2}$ & 1.127 \\
a (AU)  & 1.77 $\pm$ 0.04  \\            % inserts double horizontal lines
e  & 9.95 $\times 10^{-2}$ $\pm$ 0.32 $\times 10^{-2}$ \\                
i (degrees)  & 27.43 $\pm$ 4.40  \\    %inserts single line
$\Omega$ (degrees) & 134.95 $\pm$ 0.40 \\
$\omega$ (degrees)  & 100 $\pm$ 10 \\
$T_{\mathrm{p}}$ (years)  & 2006.039  $\pm$ 0.010 \\
T (years)  & 5.15 $\pm$ 0.09\\
M ($M_{Jup}$) & 220 $\pm$ 21\\
\noalign{\smallskip}
\hline
\end{tabular}
\end{table}

\subsection{Photometry}

Table~\ref{tab:table3} summarizes the magnitude differences between
TWA22 A and B, measured with NACO at different wavelengths. Taking
into account the filter tranformations between NACO and 2MASS and the
photometry of the unresolved system given from the 2MASS Survey
\citep{2003tmc..book.....C}, we derived the apparent JHK magnitudes of
each component (see Table~\ref{tab:table4}). Observations under bad
seeing conditions were excluded. Based on an accurate distance (17.53
$\pm$ 0.21 pc) determination \citep{2009Teixeira}, the absolute
magnitudes were also derived.  TWA22 being a young mid-M system, we monitored its photometric variations in the H band. We only noticed a 0.05 variations of the total magnitude of the system along time. This variation is reported in the error bars on our photometry in  Table~\ref{tab:table4}.    

%However, we didn't noticed photometric variations between the 2MASS and the DENIS \citep{1994ExA.....3...73E, 2005yCat....102002T} near-infrared photometry of the system despite the 1 year time-lag between DENIS and 2MASS observations (Magnitudes and colors were converted using \cite{2001AJ....121.2851C}).} 

\begin{figure}[t]
   \centering
   \includegraphics[width=\columnwidth]{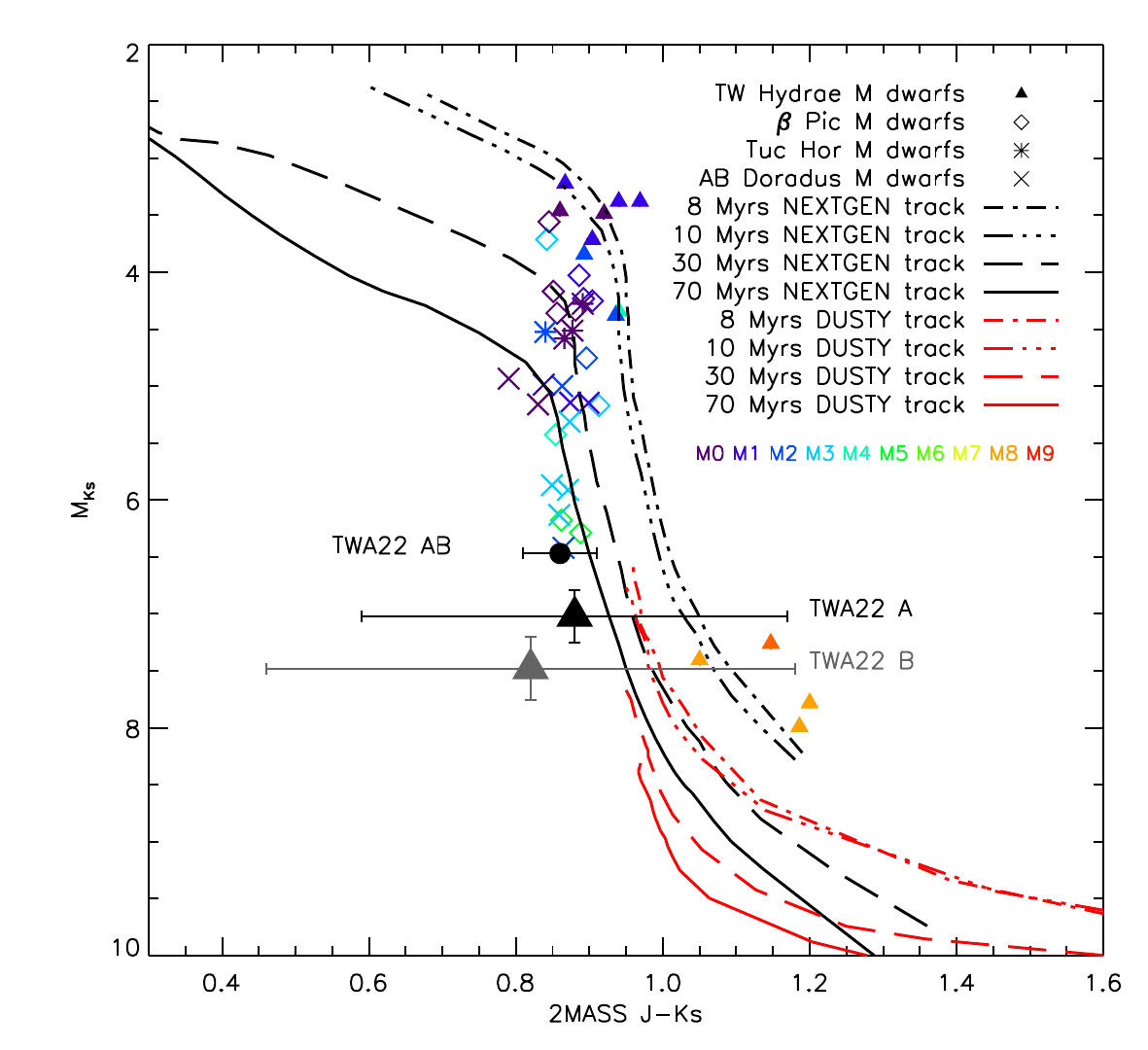}
      \caption{Color ($J-K$) - magnitude ($M_{Ks}$) diagram of
      TWA22\,A and B compared with the photometry of young M dwarfs
      members of the TW Hydrae (age=8 Myr, filled triangles), $\beta$
      Pictoris (age=10 Myr, diamonds), Tuc-Hor (age=30 Myr, stars) and AB
      Dor (age=70 Myr, crosses) associations. Distances and spectral
      types were taken from \cite{2008arXiv0808.3362T} exepted for
      the members of the TW Hydrae moving group
      \citep{2005ApJ...634.1385M}. Typical error bars are 0.03 for
      $M_{K}$ and 0.08 for J-K$_{s}$. NEXTGEN  \citep
      {1998A&A...337..403B} 8 Myr (dashed-dotted black line), 10 Myr
      (long-dashed-dotted black line), 30 Myr (long-dashed black line) and 70 Myr isochrones
      (full black line) are overplotted. DUSTY tracks \citep{2000ApJ...542..464C} at identical ages are added (red lines). These tracks were shifted by 0.2 mag to redder J-Ks. This artificially compensates the remaining incompleteness, at relatively warm temperatures, in the AMES linelists used in DUSTY  atmospheric models \citep{2007ApJ...657.1064M}.   The NEXTGEN tracks also appears bluer than 10 Myrs old mid-M  dwarfs by $\sim$ 0.2 mag, which might be related to a non full representation of their spectral energy distribution.  At a given age, early-type M dwarfs
      (violet to green points) are bluer and more luminous than
      late-type M dwarfs (green to red points) in the K band. At a
      given spectral type (i.e. temperature) the objects go fainter
      with age. The photometry is compatible with the one of young
      mid-M dwarfs. The unresolved 2MASS photometry is plotted for comparison (black dot).}
         \label{M_K_vs_JK_2MASS}
   \end{figure}

Reported in a color ($J-K$) -magnitude ($M_{Ks}$) diagram, the TWA22\,A and B photometry can be
compared with the photometry of M dwarfs of the young, nearby
associations TW Hydrae ($\sim8$~Myr), $\beta$ Pictoris ($\sim12$~Myr),
Tucana-Horologium ($\sim30$~Myr) and AB\,Doradus
($\sim70$~Myr). Predictions of evolutionary models of \cite[also named NEXTGEN]{1998A&A...337..403B} are also given at these young ages (see Fig. \ref{M_K_vs_JK_2MASS}). Although age-dependent, the near-infrared
photometry of TWA22\,A and B appears compatible for both components
with that expected for young mid-M dwarfs but does not allow to give an age estimation for the binary. The NEXTGEN tracks also appears bluer than 10 Myrs old mid-M  dwarfs by $\sim$ 0.2 mag, which might be related to a non full representation of their spectral energy distribution. We provide in the following an improved estimation of the spectral type of our targets, using our spectroscopic data.

%Finally, 2MASS
%magnitudes of TWA22\,AB were converted to MKO magnitudes using the
%\cite{2006MNRAS.373..781L} formulaes. Based on a bolometric correction
%of \cite{2004AJ....127.3516G} for a M6 $\pm$ 2 spectral type,} we infer
%a luminosity of log(L/L$_{\odot}$) = -2.11$ \pm 0.17$ for TWA22 A and
%log(L/L$_{\odot}$) = -2.30$ \pm 0.17$ for TWA22 B. \textbf{!!Corrigé ici les valeurs de luminosite!!}

\begin{table}[t]
\caption{TWA22 A and B individual magnitudes converted into the 2MASS system.}             % title of Table
\label{tab:table4}
\centering
\begin{tabular*}{\columnwidth}{@{\excs}lllll}     % 7 columns
\hline \hline                  % inserts double horizontal lines
Band & $m_{A}$ &   $m_{B}$ &  $M_{A}$ &   $M_{B}$ \\
     & (mag)                 & (mag)                   & (mag)                 & (mag)                  \\
\hline 
$J$ &  9.12 $\pm$ 0.10 & 9.52 $\pm$ 0.11 & 7.90 $\pm$ 0.13 & 8.30 $\pm$ 0.14 \\
$H$ &  8.61 $\pm$ 0.15 & 9.12 $\pm$ 0.15 &  7.39 $\pm$ 0.18 & 7.90 $\pm$ 0.18 \\
$K_{s}$ &  8.20 $\pm$ 0.19  & 8.70 $\pm$ 0.25 &  7.02 $\pm$ 0.23 &  7.48 $\pm$ 0.28 \\
\hline                                   %inserts single line
\end{tabular*}
\end{table}

% \begin{figure}
%   \centering
%   \includegraphics[width=\columnwidth]{JH_vs_HKs_2MASS.pdf}
%      \caption{Color-color magnitude diagram with indicated spectra types. TWA22 AB are overlapping the young ($\sim$5 Myr) Upper Sco M dwarfs identified by \cite{2006AJ....131.3016S}. The colors of these objects were deredden with a interstellar extinction law \citep{1985ApJ...288..618R}. The 8 myr evolutionnary track of \cite{1998A&A...337..403B} are traced for comparison. TWA22 AB agree within 1 $\sigma$ to the tracks.}
%         \label{JH_vs_HKs_2MASS}
%   \end{figure}

% \begin{figure}
%   \centering
%   \includegraphics[width=\columnwidth]{JK_vs_SpType_2MASS.pdf}
%      \caption{J-K ploted against spectral type. TWA22 AB are overlapping the young ($\sim$5 Myr) Upper Sco objects}
%         \label{JK_vs_SpType_2MASS}
%   \end{figure}

\subsection{Spectroscopic Analysis}

\subsubsection{Line identification}

\begin{figure}
   \centering
   \includegraphics[width=\columnwidth]{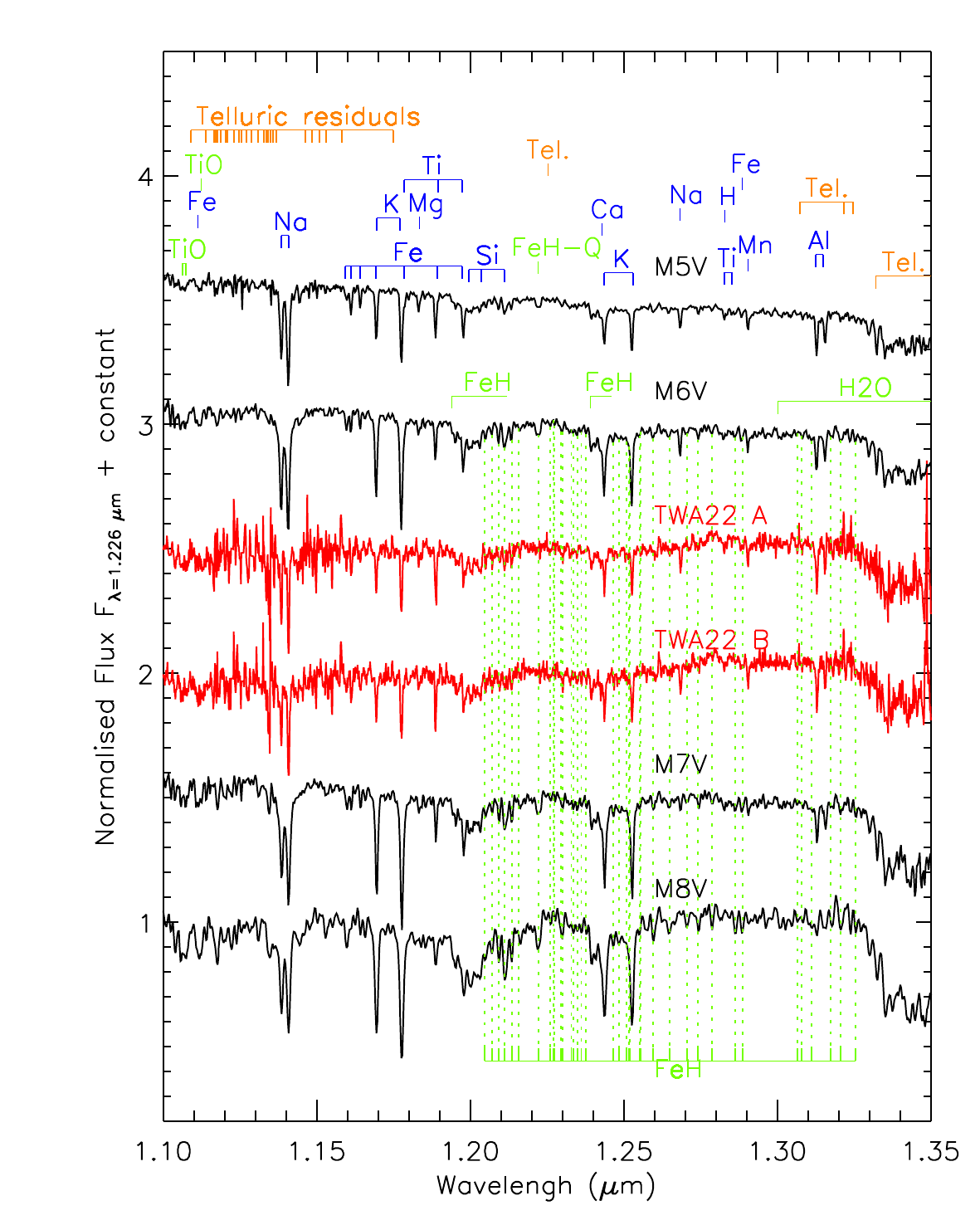}
\caption{TWA22 A and B spectra compared to spectra of M5 to M8 dwarfs. The M6 dwarf spectrum reproduces well the J bands of TWA22 A and B. However, our spectra seem to have a slightly redder slope. We reported identified atomic features (blue). Molecular absorptions (FeH bands were identified by \cite{Cushing}) are flagged in green and telluric residuals in orange. Atomic absorptions are indicated in blue.} 
         \label{Plot_Spectral_identification_Twa22A_J}
   \end{figure}

\begin{figure}
   \centering
   \includegraphics[width=\columnwidth]{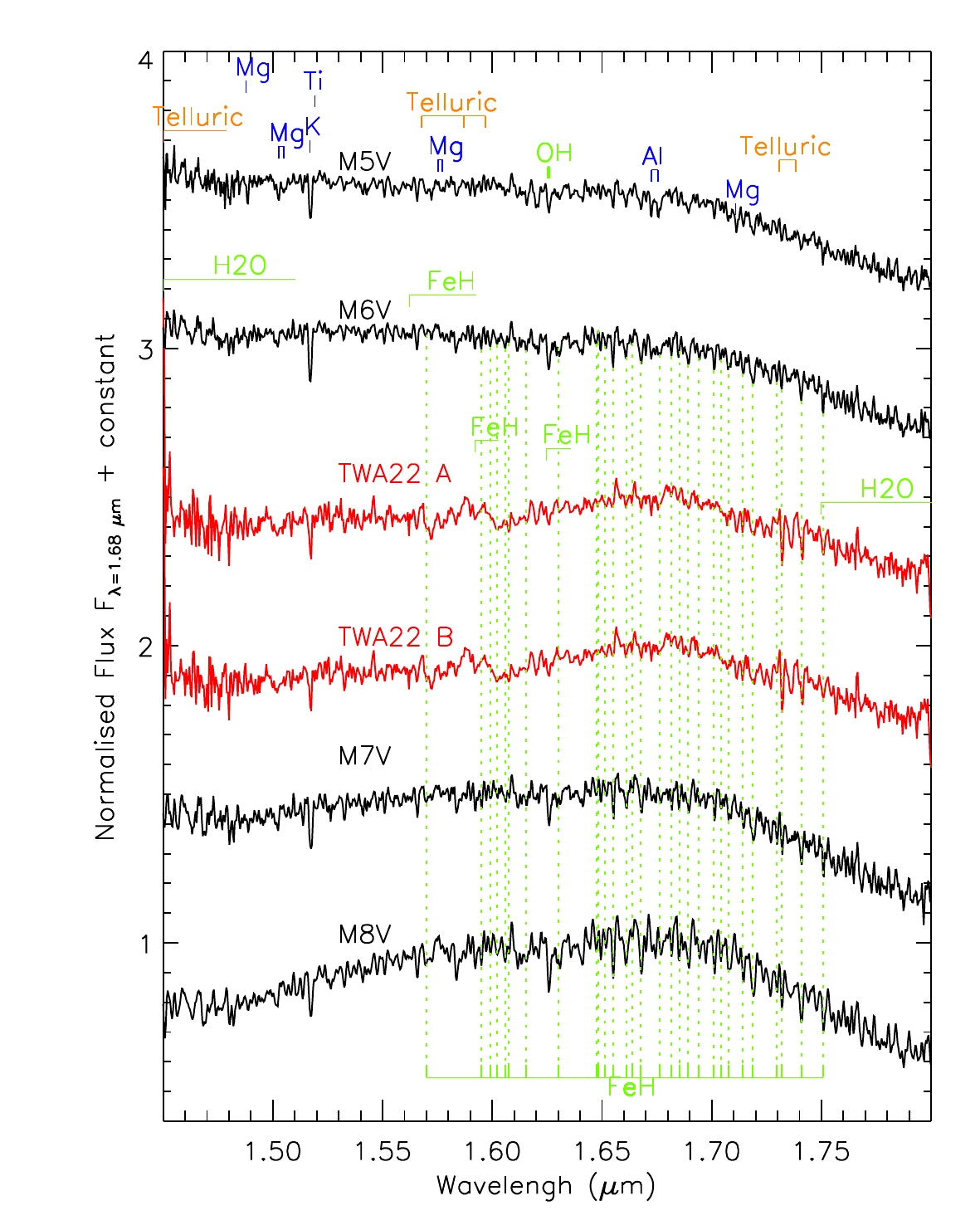}
      \caption{Same as Fig. \ref{Plot_Spectral_identification_Twa22A_J} but for H band. In this case, differences between M6V field dwarf spectrum and TWA22 spectra are important. This could arise from low gravity or flux losses introduced either in the standard star datacube and during the extraction process. } 
         \label{Plot_Spectral_identification_Twa22A_H}
   \end{figure}

\begin{figure}
   \centering 
   \includegraphics[width=\columnwidth]{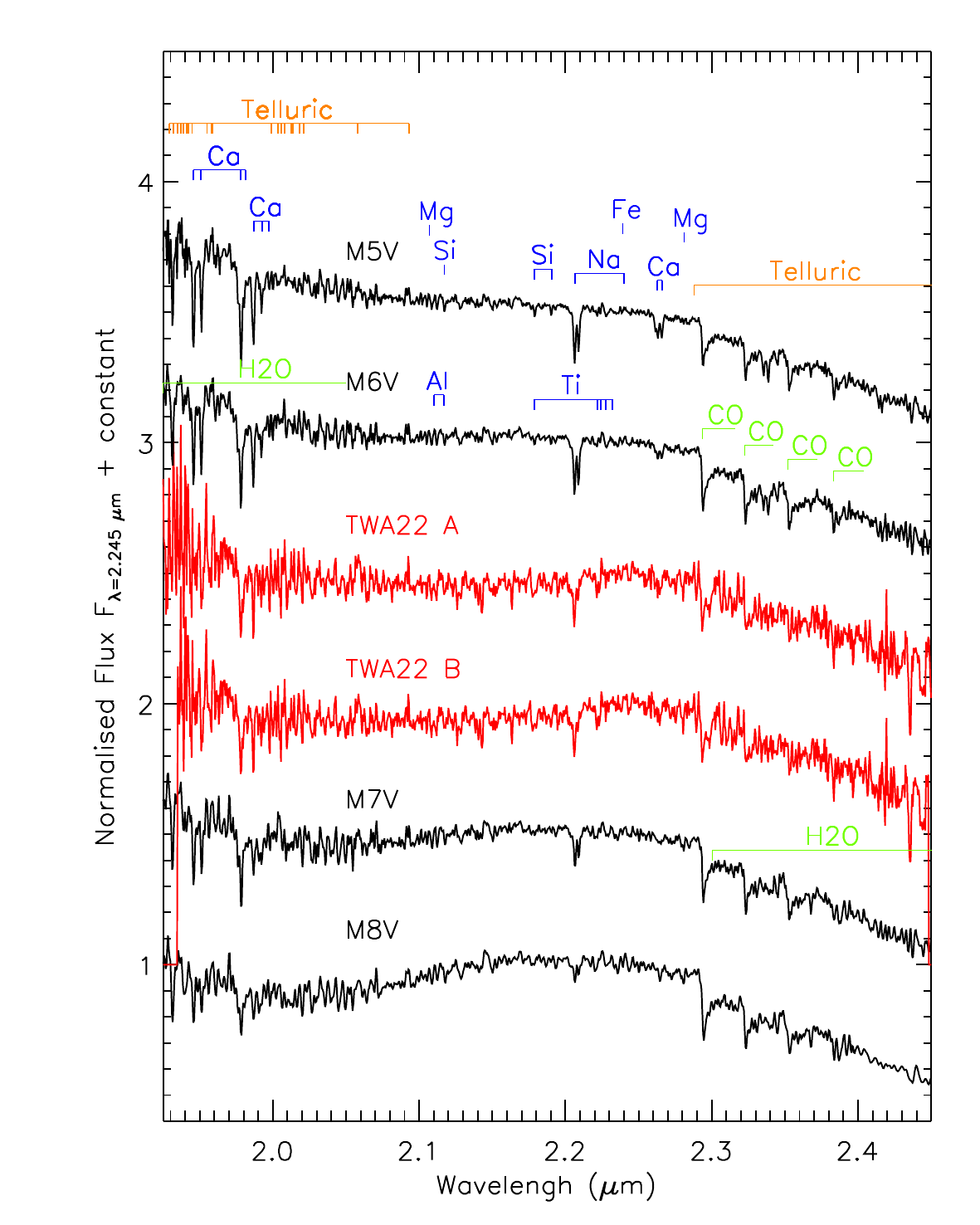}
      \caption{Same  as Fig. \ref{Plot_Spectral_identification_Twa22A_J} but for the K band. Our spectra are still contaminated by 50 Hz pick-up noise residuals. This effects is noticible  around 2.3 $\mu$m. Spectra look like M6V field dwarf.} 
         \label{Plot_Spectral_identification_Twa22A_K}
   \end{figure}

To identify the numerous spectral features in the TWA22\,A and B
spectra between 1.10 to 2.45~$\mu$m, the spectra were compared with an homogeneous
medium resolution ($R_{\lambda}\sim2000$) sequence of field dwarfs from \cite{2005ApJ...623.1115C} (hereafter C05; see Fig.
\ref{Plot_Spectral_identification_Twa22A_J},
\ref{Plot_Spectral_identification_Twa22A_H} and
\ref{Plot_Spectral_identification_Twa22A_K}). The TWA22\,A and
B spectra appear very similar.

In \textit{J}-band, both TWA22\,A and B spectra are dominated by the strong Na
I doublet at 1.138 $\mu$m, the deep K I lines at 1.169, 1.177, 1.243
and 1.253 $\mu$m and the presence of a broad H$_{2}$O absorption from
1.32 to 1.35 $\mu$m.  Fe I absorptions are also detected around 1.170
$\mu$m. One is blended with the 1.177 $\mu$m K I line. We notice additional broad
FeH absorptions around 1.20 $\mu$m and 1.24 $\mu$m compatible with
that expected for mid-M dwarfs as well as the presence of the weak Q-branch
at 1.22 $\mu$m. Finally, the Al I doublet at 1.313 $\mu$m is
detected. This doublet is expected to disappear at the M-L transition.

In \textit{H}-band, the spectra are affected by H$_{2}$O absorptions from 1.45
to 1.52 $\mu$m and from 1.75 to 1.8 $\mu$m. They exhibit pronounced K
I atomic lines at 1.517 $\mu$m as well as weak doublets of Mg I at
1.503 $\mu$m and Al I at 1.675 $\mu$m. Weak FeH absorptions are also
present. They increase from M5 to the M-L transition
\cite{Cushing} and their depths are here compatible with those expected for
M5 to M7 field dwarfs.

In \textit{K}-band, strong H$_{2}$O absorptions appear from 1.95 to 2.04 $\mu$m
and from 2.3 to 2.45 $\mu$m. They are typical from mid-M to mid-L dwarfs. Strong
Ca I features are present from 1.9 to 2.0 $\mu$m. They tend to disappear
in the spectra of field dwarfs at the M-L transition. We identify
firmly the first overtone of CO near 2.3 $\mu$m, the rest being
affected by the 50~Hz pick-up noise oscillations mentionned earlier. Additional weak Mn I,
Ti I, Mg I and Si I absorptions are spread over the J, H and K
bands. These lines are expected to be rapidly replaced by molecular
absorptions for dwarfs later than M5. The 1.106 $\mu$m band seems to
be overlapping H$_{2}$O and TiO absorptions with increasing depths
from early to late M dwarfs. Finally, the 1.626 $\mu$m feature
corresponds to close OH lines, as noted in \cite{1996ApJS..104..117L}.

To conclude, the features detected over the spectra of TWA22\,A and B
between 1.1 and 2.45~$\mu$m suggest that both components have a cool
atmosphere, typical of mid to late-M dwarfs.

\begin{figure}
   \centering
   \includegraphics[width=\columnwidth]{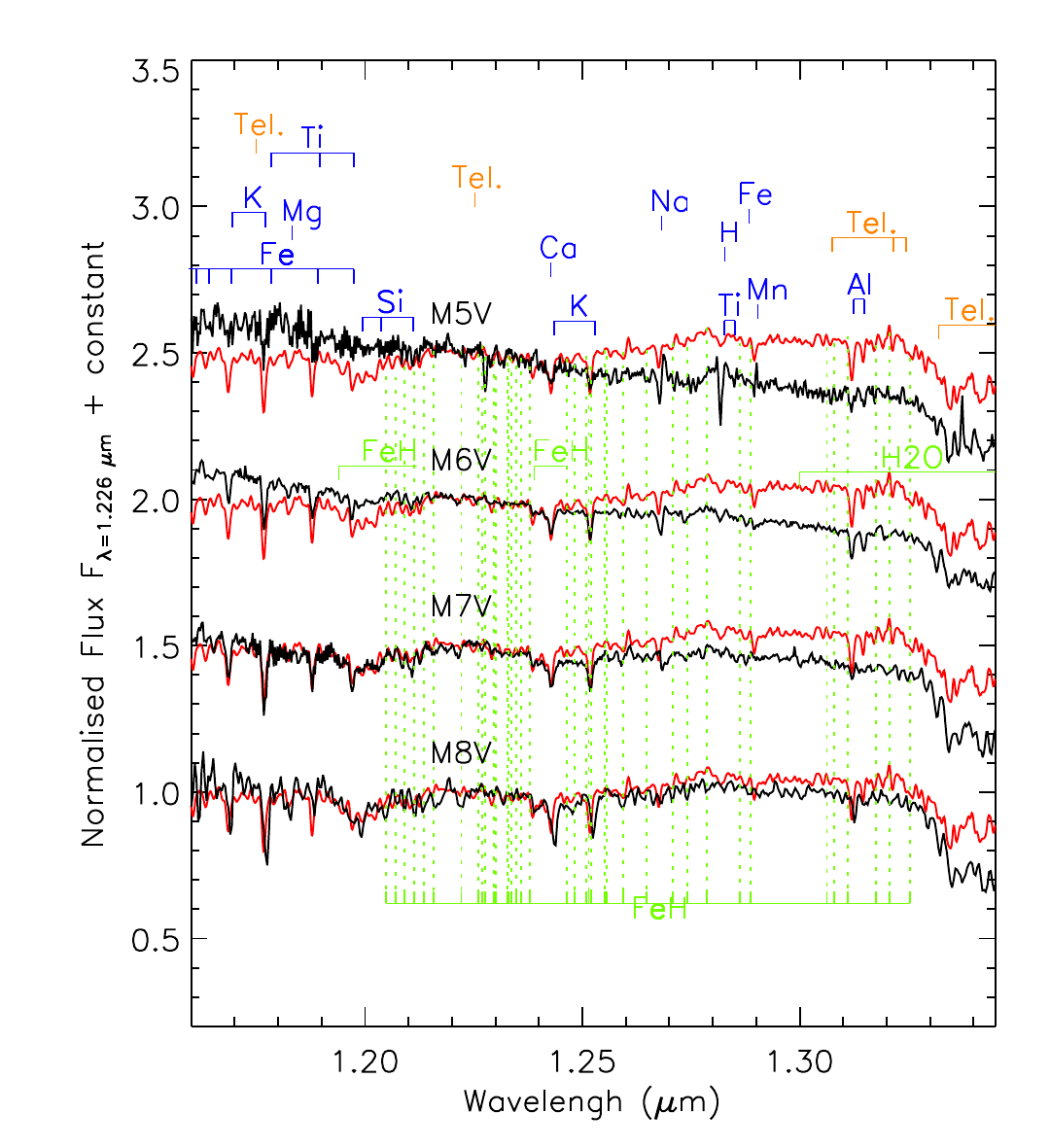}
\caption{Comparison of the TWA22 B J band spectrum (red) to spectra (black) of young Upper Sco and Orion nebulae cluster objects \citep{2004ApJ...610.1045S, 2008MNRAS.383.1385L}. We clearly notice that the TWA22 B spectral slope is redder than that of reference spectra. The TWA22 B spectrum is very similar to those of young M6 and M7 dwarfs. We reported identified atomic features (blue). Molecular absorptions (FeH bands were identified by \cite{Cushing}) are flagged in green and telluric residuals in orange. Atomic absorptions are indicated in blue.} 
         \label{Plot_Spectral_identification_Twa22A_J_young}
   \end{figure}

\begin{figure}
   \centering
   \includegraphics[width=\columnwidth]{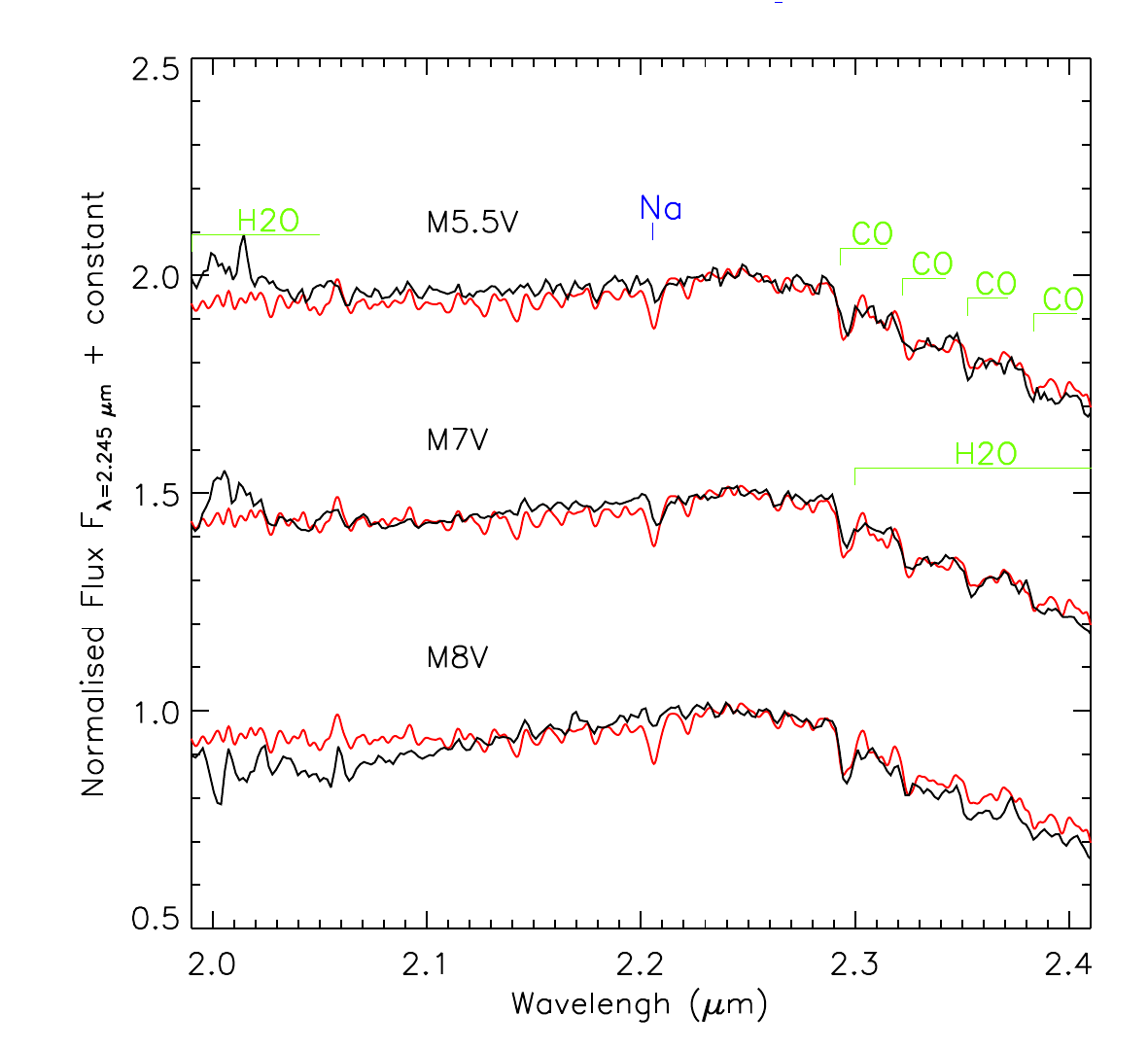}
      \caption{Comparison of the TWA22 B K band spectrum (red) to spectra of young Upper Sco standards (black) at R=350 \citep{2003ApJ...593.1074G}. The TWA22 B spectrum was convolved with a gaussian to match the resolution of standard star spectra. Our spectrum is reproduced by spectra of young  M5.5 and M7 dwarf.} 
         \label{Plot_Spectral_identification_Twa22A_K_gorl}
   \end{figure}

\subsubsection{Continuum fitting and spectral indexes}

The continuum of both TWA22 A and B spectra were compared to spectra
of field M dwarfs obtained by C05 and \cite{Mclean2003}, hereafter
ML03. Least squares were computed on parts of the spectra free from telluric
correction residuals. From 1.10 $\mu$m to 1.27 $\mu$m, the TWA22 A and
B continuums are well reproduced by M6 $\pm$ 1 dwarfs. The \textit{H}-band
spectra are visually poorly reproduced. Least squares are minimized
for M9 dwarfs but with 2 subclasses uncertainty. Finally, our \textit{K}-band
spectra are well fitted by M5 to M7 dwarfs. From these comparisons, we
assign a spectral type M6 $\pm$ 1 to both TWA22\,AB components.

TWA22\,AB being a young system, we have tested if using high surface
gravity spectra of old field dwarfs might affect our spectral
analysis. Intermediate surface gravity reduces the strength of
alkali lines \citep{2001MNRAS.326..695L,2003ApJ...593.1074G,
2004ApJ...600.1020M, 2006ApJ...639.1120K} and produces triangular
shape in \textit{H}-band interpreted as collision induced absorptions (CIA) of
H$_{2}$. Our spectra were then compared with young (age $\lesssim$ 8 Myr) dwarfs
spectra \citep{2003ApJ...593.1074G, 2004ApJ...610.1045S, 2008MNRAS.383.1385L} in the \textit{J} and \textit{K} bands (see Fig. \ref{Plot_Spectral_identification_Twa22A_J_young} and \ref{Plot_Spectral_identification_Twa22A_K_gorl}). They are mostly similar to M5, M6 and M7 dwarf spectra, and consistent with the continuum fit obtained with field dwarfs. In both cases, our J-band spectra are slightly redder and our H-band spectra visually are still poorly reproduced by young and old M dwarfs.

\begin{figure}
   \centering
   \includegraphics[width=\columnwidth]{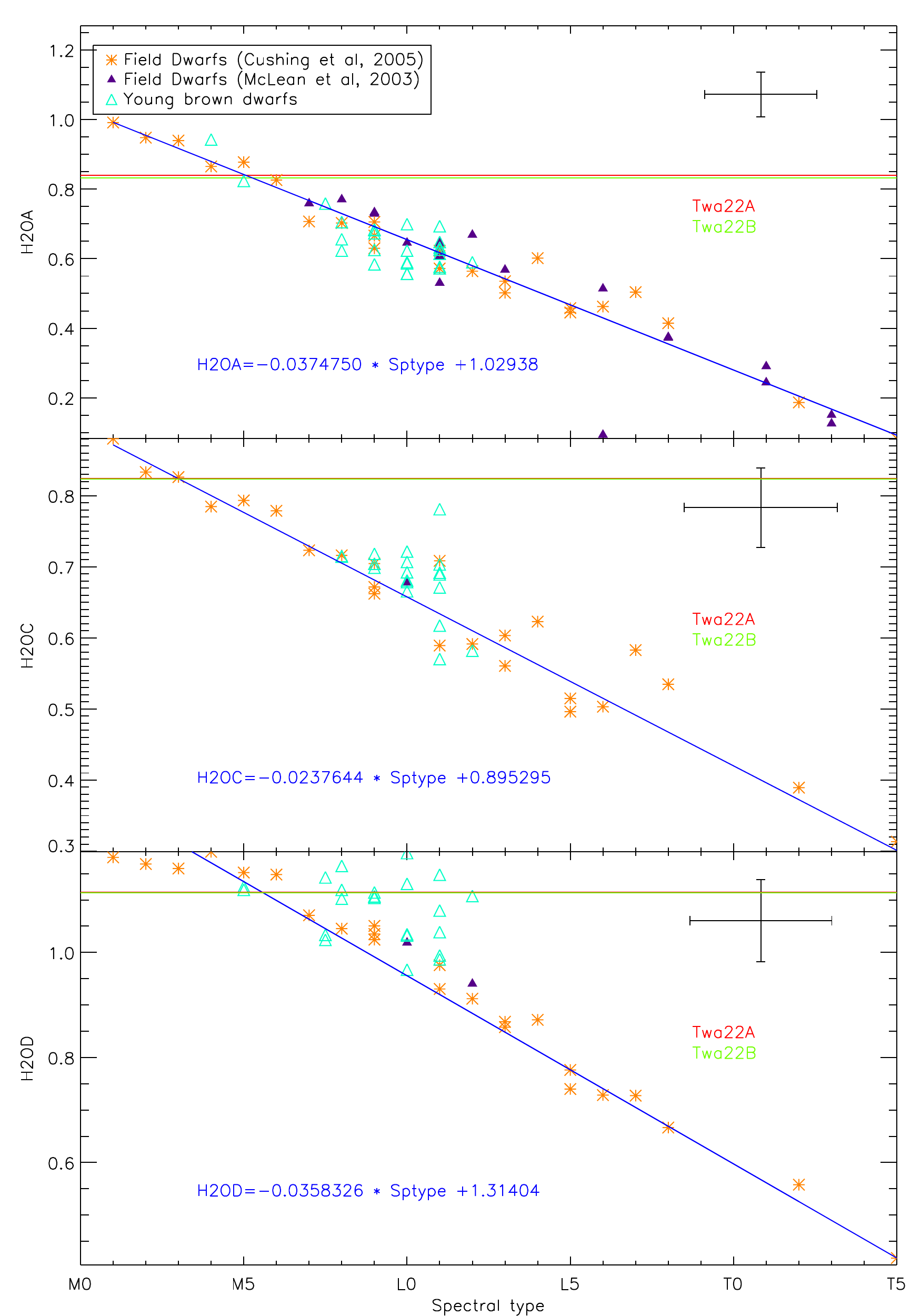}

      \caption{H$_{2}$O spectral indexes computed on libraries of
      field \citep{Mclean2003, 2005ApJ...623.1115C} and young dwarfs spectra \citep{2004ApJ...610.1045S, 2008MNRAS.383.1385L}. Redundancies between
      libraries have been checked. Young dwarfs M4-L2 dwarfs follow
      trends of field dwarfs exept for H$_{2}$O D. The dispersion of
      values for young upper Sco brown dwarfs does not seem to arise from
      redenning effects. When overplotted, TWA22 A and B values point
      an M6 $\pm$ 1 spectral type. In the upper right corner of each plot are reported the $1\sigma$ errors derived from the residuals to the linear fit.}

         \label{Plots_H2Oindex_Twa22AB}
   \end{figure}

To complete this spectral type determination, spectral indexes
developed by ML03 (from $H_{2}O$ bands at 1.34 $\mu$m ($H_{2}OA$),
1.79 $\mu$m ($H_{2}OC$), 1.96 $\mu$m ($H_{2}OD$) and at 1.2 $\mu$m
from the FeH band) were derived for TWA22\,A and B (see Fig. \ref{Plots_H2Oindex_Twa22AB}). The results were
compared to the values computed from the ML03 and C05 spectral libraries
of field dwarfs. They were also compared to values derived for 
young dwarfs (\cite{2004ApJ...610.1045S, 2008MNRAS.383.1385L},
hereafter S04 and L08) to test the sensitivity of these indexes to surface
gravity (age). In fact, The $H_{2}O$ D and FeH indexes values tend to
increase with age for M5-L2 dwarfs, and could disturb our analysis. We
then used a mean weight of the individual spectral type estimations
from $H_{2}OA$, $H_{2}OC$ and the recently defined Allers $H_{2}O$
index at 1.55 $\mu$m (see \cite{2007ApJ...657..511A}) to infer $M5\pm
1$ and $M5.1\pm 1$ spectral types for TWA22 A and B
respectively. These results mach the M6 $\pm$ 1.5 and M5 $\pm$ 1
values derived for TWA22\,A and B from the $H_{2}OD$ and FeH indexes
for the 2 objects.  Based on the \textit{K}-band photometry and the
associated bolometric corrections of \cite{2004AJ....127.3516G}, we
derive a luminosity of \textit{log(L/L$_{\odot}$)}=-2.11$ \pm 0.13$
dex for TWA22 A and \textit{log(L/L$_{\odot}$)}=-2.30$ \pm 0.16$ dex
TWA22 B. Using the T$_{eff}$-spectral type conversion scales for
intermediate-gravity objects \citep{2003ApJ...593.1093L}, we find an
initial estimation of $T_{eff}=2990^{+135}_{-190}$ K for both
components.

\subsubsection{Study of narrow lines}

Depths of many narrow lines were studied to provide additional information
on the surface gravity of TWA22\,AB and particularly
on its age. Following the \cite{1992ApJS...83..147S} method to
measure pseudo-equivalent widths and their associated uncertainties, we derived
the equivalent widths of strong atomic lines over the
\textit{J}, \textit{H} and \textit{K} bands. They were computed for narrow lines at 1.106 $\mu$m (TiO and $H_{2}O$), 1.220 $\mu$m (FeH - Q branch),
1.313 $\mu$m (Al I) and 1.626 $\mu$m (OH), and for
the K I doublets at 1.169, 1.177, 1.243 and 1.253 $\mu$m. The
results were compared with pseudo-equivalent widths of old
field dwarfs (C05, ML03) and young Upper Sco dwarfs (S04,
L08). The use of both librairies confirmed the strong surface
gravity dependency of the K I lines, more moderate for the
Al I, FeH and OH lines. Due to the degeneracy in terms of
effective temperature and surface gravity, pseudo-equivalent
widths alone are not sufficient for a precise spectral type determination
of TWA22 AB. They remain however compatible
with narrow lines depths of young and old dwarfs of spectral
types later than M4.

\begin{figure}
   \centering
   \includegraphics[width=\columnwidth]{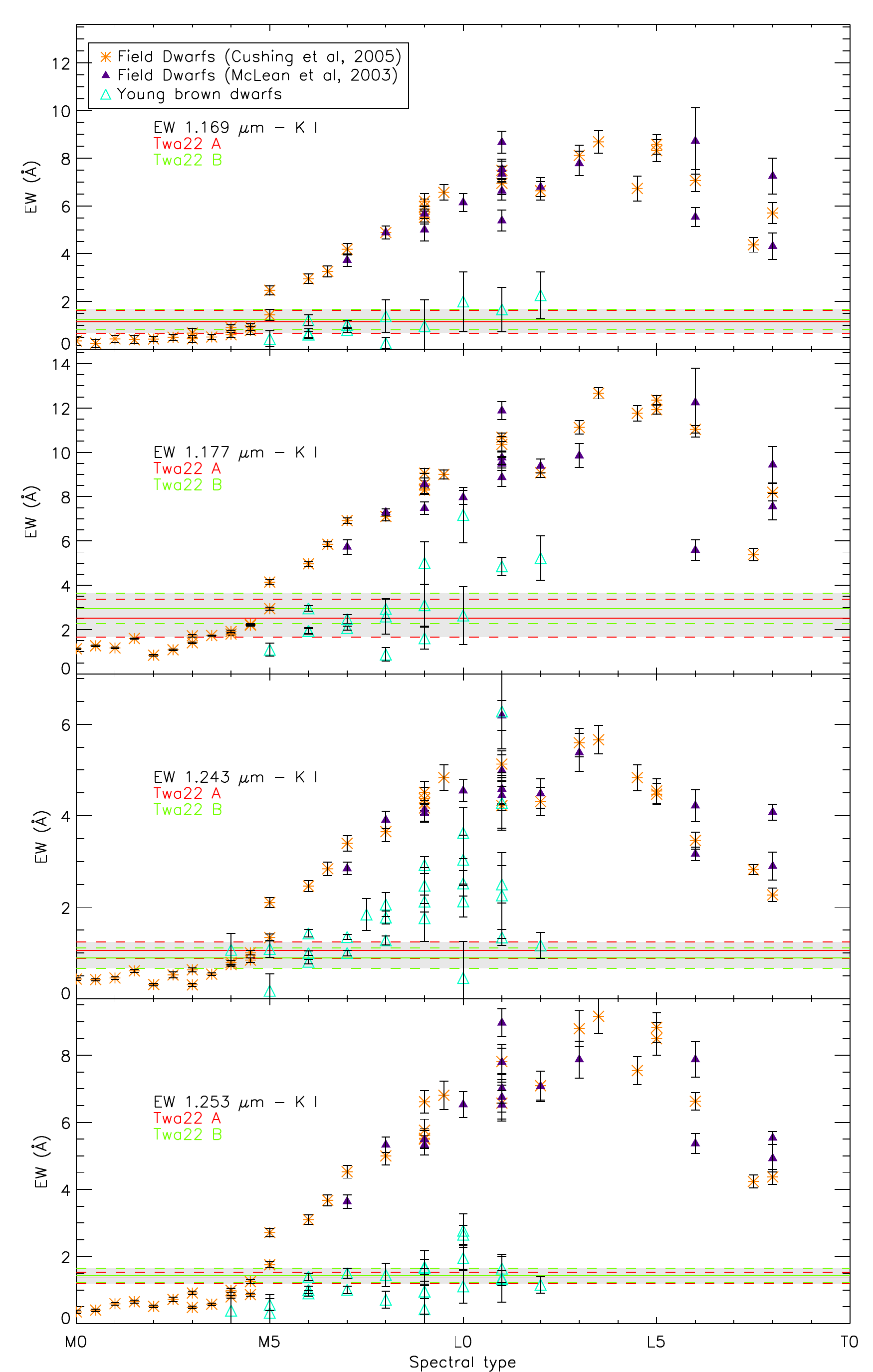}

      \caption{Equivalent widths in the K I lines computed on
      libraries of young and field dwarfs spectra. Young dwarfs spectra have
      weak K I lines and therefore low equivalent widths compared to
      field dwarfs as a consequence of their low surface gravity. TWA22 A and B values are reported as red and green lines respectively with their associated uncertainties (dashed lines). K I equivalent widths of field dwarfs point a M5V spectral type different from our M6 $\pm$ 1 spectral type estimation from continuum fitting. This effect is probably the result of weakening of K I lines associated to the estimated very young age of the system.}

         \label{Plot_Largeurs_equivalentes_Twa22_KI}
   \end{figure}

If we now assume a spectral type M6 $\pm$ 1 for both components, the
pseudo-equivalent widths of TWA22 AB appear intermediate between
values found for young and field dwarfs (see Fig.
\ref{Plot_Largeurs_equivalentes_Twa22_KI}). This is confirmed using a
visual comparison with an evolutionary sequence of M6 dwarfs composed
of the old field dwarf GL 406, the intermediate old companion AB Doc
C (Age 75 Myr, M5.5, \cite{2007ApJ...665..736C}) and a young M5.5
dwarf from the Orion nebulae (age $\sim$ 1-2 Myr, S04). Together with
the other age indicators, these intermediate surface gravity features
confirm that TWA22 AB is likely to be a young binary system.  However,
their uncertainties remain significantly large to not assign a precise
age.

%\begin{figure}
%   \centering
%   \includegraphics[width=\columnwidth]{Plot_Largeurs_equivalentes_Twa22_multi.pdf}
%
%      \caption{Evolution of equivalent widths of the 1.106 (TiO and
%      $H_{2}O$), 1.220 (FeH), 1.313 (Al I) and 1.626 $\mu$m (OH)
%      features with spectral type. The TiO line can be used to assign
%      spectral type. Other lines can be used to asign an M6 spectral
%      type to TWA22 A and B.}
%
%         \label{Plot_Largeurs_equivalentes_Twa22_multi}
%   \end{figure}

\subsection{Gravity and effective temperature from atmospheric models}

For a fine determination of the effective temperatures and surface
gravities of TWA22\,A and B, we compared our observed spectra with
theoretical templates from the GAIA model v2.6.1
\citep{2005ESASP.576..565B}. This library is updated from
\cite{Allard_2001}. It benefits from improved molecular dissociation
constants, additional dust species with opacities, spherical symmetry,
and a mixing length parameter 2.0 $\times$ H$_{p}$. The temperature ranges in
the templates from 2000 to 10 000 K and the gravity from -0.5 to 5.5
but we limited our analysis to $ 2000 K \leq T_{eff} \leq 4000 K $ and
$3.5 \leq log(g) \leq 5.5 $. Theoretical spectra were convolved with a
gaussian to match the SINFONI spectral resolution and interpolated to
the TWA22\,AB wavelength grid. Least squares minimization was applied
to find templates that fit the TWA22 A and B continuum avoiding zones
polluted by remaining oscillations.

The TWA22 A least-square map in the J band constrains the temperature
between 2800 to 3100 K and is minimized for log(g)=4.5 and
$T_{eff}$=2900 K. 
%We added a blue slope to the TWA22 A J band in order
%to simulate the impact of the differential flux losses. A minimum
%least square value is found for 20 \% flux losses and shift the
%estimated TWA22 A temperature to 3000 K. 
In \textit{H+K} band, our
minimization failed to reproduce faithfuly the TWA22\,A spectra and
makes us suspect the existence of a constant flux loss in H band
during the spectral extraction process. To limit this systematic
effect, the minimization was applied separately in \textit{H} and
\textit{K} bands. In \textit{H}-band, the effective temperature is
minimized between 2600 K to 3000 K in the full space of surface
gravities explored. The minimum is located at 2800 K and
log(g)=4.5. The \textit{K} band is well reproduced by 2900 and 3000 K
templates irrespective of gravity. Summing the three bands, we
estimate an effective temperature $T_{eff}$=2900$^{+200}_{-200}$ K for
TWA22 A. Conducting a similar analysis for the component TWA22 B, we derive an
effective temperature $T_{eff}$=2900 $^{+200}_{-100}$ K. Using the \cite{2003ApJ...593.1093L} scale, these temperatures estimations correspond respectively to M7$^{+1}_{-2}$ and M7$^{+0.5}_{-2}$ spectral types for TWA22 A and B. This is also consistent with spectral types estimated in part. 4.3.2.

\begin{figure}
   \centering
   \includegraphics[width=\columnwidth]{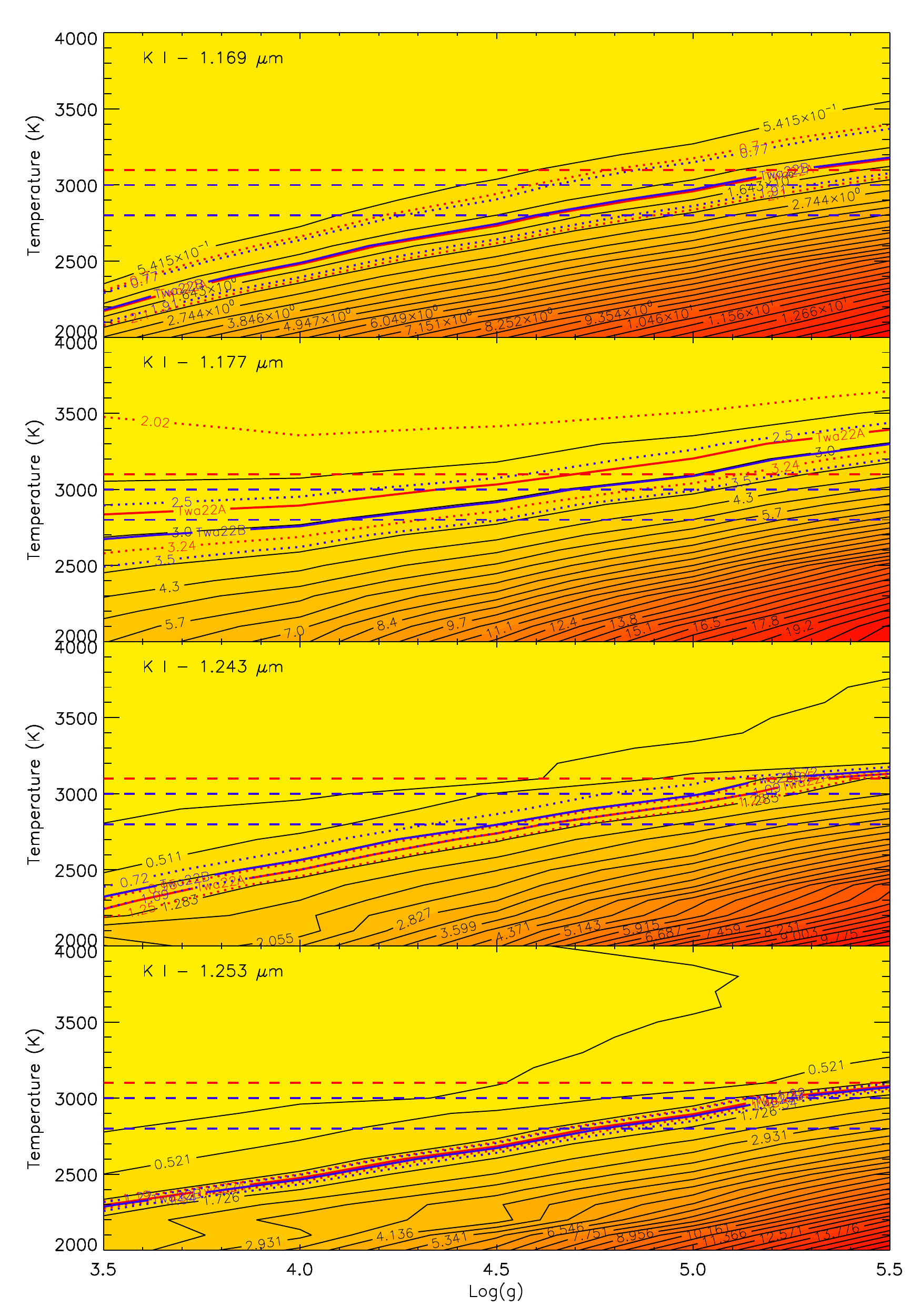}

      \caption{Iso-contours plots of K I lines equivalent widths computed
 on each spectral templates of the GAIA library v2.6.1 Red
 color indicates high values. The contours for TWA22 A (red) and B
      (blue) pseudo-equivalents widths values are overplotted. The
 long-dashed lines represent limits on TWA22 A (red) and TWA22 B
      (blue) temperatures. Gravity is estimated inside these
 temperatures boxes.}

         \label{EW_KI_vs_temperature_vs_gravity_contour}
   \end{figure}

For a fine determination of the surface gravity from synthetic
spectra, we computed the equivalent widths of K I lines in the
J band on each spectra of TWA22\,A and B. We then compared
the values to TWA22 A and B to restrain the acceptable gravity domain
(see Fig. \ref{EW_KI_vs_temperature_vs_gravity_contour}). We then
estimate that the surface gravity is located between log(g)=4.0 and
5.5 for TWA22 A and B.

%__________________________________________________________________

\section{Discussion}

\subsection{Evolutionary models predictions}
\begin{figure}
   \centering
   \includegraphics[width=\columnwidth]{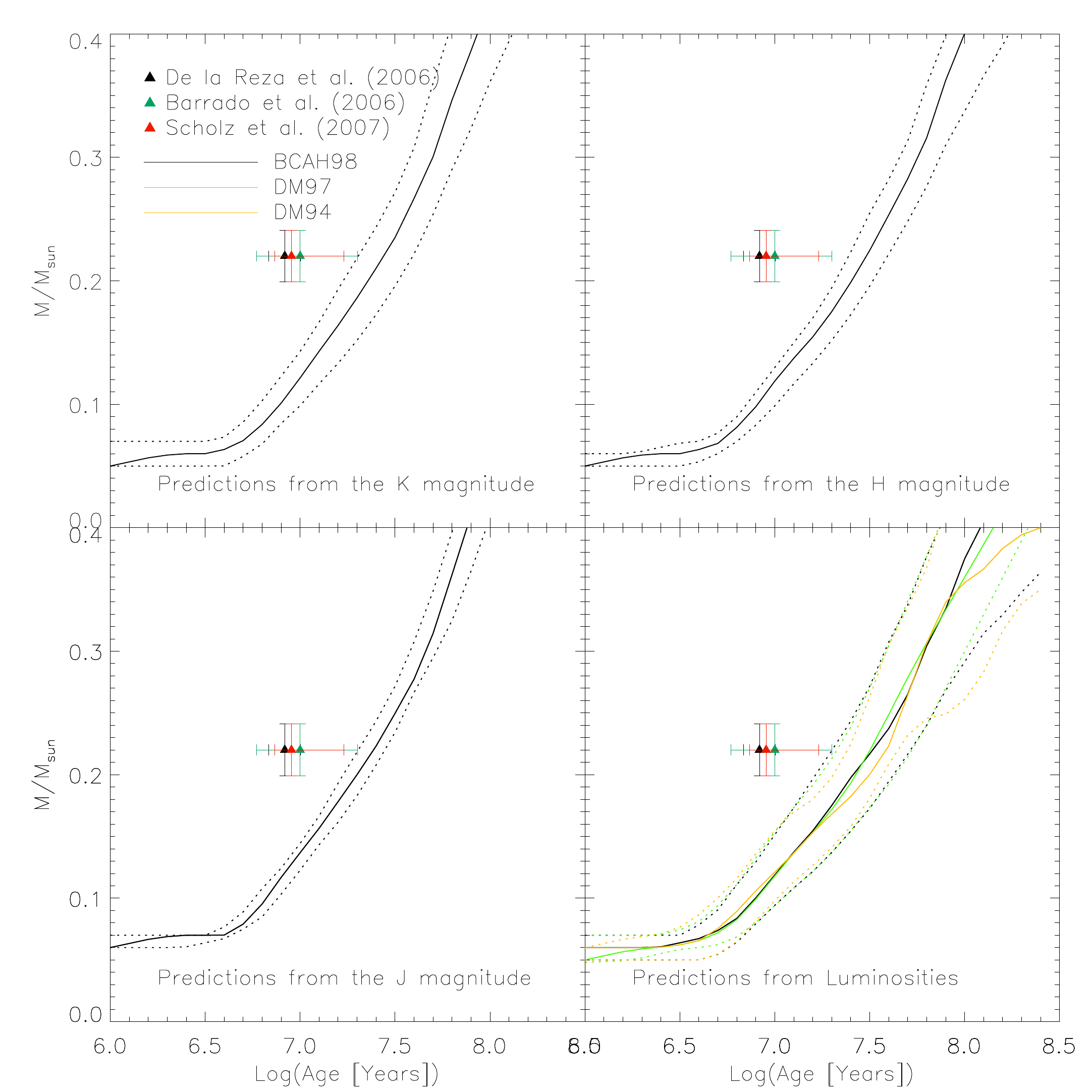}

 \caption{\textbf{Top-Left} : Confrontation of the binary direct
 mass measurement to predictions of the BCAH98 masses for different
 ages from $M_{K}$. Different estimations of the age of TW Hydrae
 \citep{2006AJ....131.2609D, 2006AA...459..511B, 2007ApJ...662.1254S}
 are reported on the graph. Errors on the photometry are propagated on
 predictions (dotted lines). \textbf{Bottom-Left} : Same as top-left
 but for predictions from $M_{H}$. \textbf{Top-right} : Same as
 Top-Left but for predictions from $M_{J}$. \textbf{Bottom-right} :
 Same as Top-Left but for predictions from our estimated TWA22 A/B
 luminosities. In this plan, predictions from BCAH98, DM94 and DM97
 models are nearly the same within our uncertainties.}

         \label{mass_vs_age}
   \end{figure}

The membership of TWA22\,AB to TW Hydrae constrains the age of the system to 3-20
Myr \citep{2006AA...459..511B,2007ApJ...662.1254S,
2006AJ....131.2609D}.  Based on our astrometric observations combined
with an accurate distance determination, we were able to derive the
dynamical mass of this tight binary. From photometry and spectroscopy,
we derived near-IR fluxes, luminosity, spectral type, effective
temperatures and the surface gravity of each component. Finally,
spectroscopy tends to indicate that both components have intermediate
surface gravity features in their spectra, supporting a young age for
TWA22\,AB. Assuming the TWA age for this system, we can now compare
the measured total dynamical mass of the binary with the total mass
predicted by evolutionary models of Baraffe et al. (1998; hereafter
BCAH98). Model predictions are based on the JHK photometry, the
luminosity and the effective temperature of both components (see Fig
\ref{mass_vs_age}). At 8~Myr, BCAH98 models systematically
under-estimate the total mass by a factor of $\sim2$.  This factor varies from 3 to 1.3 between 3-5~Myr and 20~Myr. The mass is still strongly under-estimated
using other evolutionary models of very low mass stars
\citep{1994ApJS...90..467D,1997MmSAI..68..807D}. Alternatively, if we
artificially change the system age to 30 Myr, the model predictions
match relatively well our observations.

The apparent discrepancy between our observations and the model
predictions at the age of TWA leads us to consider four explanations:
\begin{enumerate}
\item Remaining uncertainties are present in our data reduction
and interpretation related to the astrometry, photometry and
spectroscopy extraction process, the atmosphere model used or the
assumption on the system itself,
\item  The system is of higher multiplicity than observed,
\item Evolutionary model predictions are correct and the age estimate
of TWA22\,AB is currently incorrect. TWA22\,AB would be then slightly
older and aged of 30~Myr,
\item Finally, the TWA22\,AB age is 8~Myr and evolutionary models
themselves do not predict correctly the physical properties of very
low mass stars at young ages.
\end{enumerate}

Before drawing important conclusions on the validity of evolutionary
models at young ages and very low masses, we consider below the three 
first explanations.

\subsection{Data reduction and interpretation uncertainties}

Systematics on the estimation of the relative position and near-IR fluxes
of TWA22 A and B seems very unlikely. Our analysis relies on the use
of several imaging analysis techniques (aperture photometry, PSF
fitting, deconvolution), already used and tested in various
contexts. The tight binary TWA22\,AB does not represent itself a
difficult case. In addition, at each epoch, consistent results were
found on several observing sequences obtained during the
night. 

Systematics in the spectroscopic observation and extraction seem more
probable for the determination of effective temperature and surface
gravity. Differential flux losses over the $J$ or $H+K$ spectral range
may have occurred due to the limited size of the SINFONI field of view. The
impact of this effect can be simulated by adding a linear slope in our
spectral minimization over the different spectral bands. The results
do not change significantly our analysis based on continuum fitting or
spectral indexes. It does not affect at all the study of narrow lines
and our surface gravity estimation. A non-linear differential flux
loss could be responsible for our failure to faithfully reproduce the
TWA22\,A and B spectra in H-band using either empirical or synthetic
libraries. Finally, the atmosphere models were also used in various
conditions (metallicity, mixing length, different opacity tables)
without drastically changing our results.

\subsection{Higher multiplicity hypothesis}

Considering our data reduction and analysis as robust, we may wonder
whether our basic assumptions concerning the system itself are
correct. Actually, we cannot exclude from our observations that
TWA22\,AB is a multiple system of higher order. One or even both
components could be in fact unresolved binaries. In such case, the derived
effective temperatures as well as the estimated spectral types would
not be strongly modified.

Dynamically speaking, the stability of the system would require the
separation of the individual sub-components to be significantly less than
the size of the main orbit, typically by a factor 3-4 \citep{1994ApJ...421..651A}.
To refine this estimate in the present case, we performed 3-body
simulations using the symplectic code HJS \citep{2003A&A...400.1129B} dedicated to
hierarchical systems. We assume the fitted orbit and split one of
the two components into 2 equal mass bodies, with a coplanar orbit
with repect to the wide orbit and a given semi-major axis,
and assuming initial zero eccentricity. We find that the
system remains stable up to a separation of $\sim 0.4\;$AU between the two sub-components. Note that this is a priori
the most stable configuration, as a split into unequal masses would
lead to less stability for the lighter component. We also checked that
highly inclined relative configurations are physically unstable: they inevitably lead to a strong
Kozai resonance characteristic for such triple systems
\citep{1962AJ.....67..591K,1997A&A...320..478B} that cause
the eccentricity of the inner orbit to be pumped up to $\sim 1$,
leading to a physical collision between the two individual components. Hence 0.4\,AU can be considered as the widest possible separation for
hypothetical sub-components. This is in agreement with \citet{1994ApJ...421..651A}.

0.4\,AU  ($\sim$22 mas) remains below the PSF of the VLT/NACO images ($\sim 1\,$AU
given the distance of TWA22). This would explain why the inner binary
would not be resolved. However, we did not notice any PSF-lengthening
in the images. With a separation of 0.4\,AU we could expect to see
one. Does it suggest that the actual separation is significantly smaller ? 

An orbit with 0.4\,AU separation would correspond to an orbital period
of 0.8\,yr and a radial velocity wobble of $\pm
7\,\mbox{km}\,\mbox{s}^{-1}$ if we take into account the $27\degr$
inclination with respect to the plane of the sky.
Even unlikely, this modulation could not to have been detected during the monitoring (split up into two periods of 6 and 1 months). But if we assume a separation of $\sim 0.1\,$AU
to be compatible with the absence of PSF-lengthening, now the radial
velocity wobble jumps to  $\pm 14\,\mbox{km}\,\mbox{s}^{-1}$ over a
0.1\,yr period. Such a variation was not detected in the radial velocity dataset (see \cite{2009Teixeira}). Finally, no photocenter scatter is present around our two-body orbital solution. A motion of $\sim$5 mas is expected along the orbital period for a separation of $\sim$ 0.1 AU between the subcomponents. 

Ultimately, we cannot definitely rule out the possibility that at least one of
the two components of TWA22 is itself a binary, consisting of two
nearly equal mass bodies. But combined dynamical and observational
constraints show that the range of possible separations is fairly
narrow, typically 0.1--0.2 AU. Also the system needs to be at least
roughly coplanar.  
 
\subsection{Age and membership of TWA22\,AB}

Given the good agreement between observations and model predictions at
30~Myr, we can consider that the current age estimate of TWA22\,AB is
possibly incorrect. This age is currently infered from the membership
to TWA. Since the age of TWA is well established at 8 Myr from various
age diagnostics, a reliable explanation concerns the membership to TWA
itself. 

S03 identified TWA22 as a new member of TWA mainly from the
observed Li absorption line at 6708 \AA~and H$\alpha$ emission line. The Li line
is stronger (EW=510 m\AA) than those of early-M dwarfs members of
$\beta$ Pic and leads S03 to suggest an age $\leq$10 Myr (see Fig. 8 of
S03). They derived in addition a photometric distance of 22~pc for
TWA22, confirming the proximity of this young system. More recently, \cite{2005ApJ...634.1385M} discussed the membership of
TWA22\,AB to TWA based on its kinematics
properties. \cite{2005ApJ...634.1385M} estimates a probability of 2\%
for TWA22 to be a member of TWA from an implemented convergence point
technique \citep{1999MNRAS.306..381D}. However,
\cite{2006ApJ...652..724S} mentioned that the strong Li line of
TWA22\,AB is observed only for young active M dwarfs in the direction
of TWA with the exception of a very few M-type members of the $\beta$
Pictoris moving group (BPMG). Finally, \cite{2008ApJ...689.1127M} obtained a new visible high resolution spectrum of TWA22\,AB. They confirmed the strong equivalent line of the 6708 \AA~lithium absorption (EW=616 $\pm$ 21 m\AA; the strongest measured in their sample composed of young association members). They estimated Teff of 2990 $\pm$ 13 K  (compatible with the individual Teff derived in Part 4.4) and log(g)=4.2 $\pm$ 0.05 dex for the unresolved system. These new elements tend to confirm that TWA22 is a young system (age $\leq$ 30 Myr; see BCAH98 predictions).

To conciliate past and present results, we can consider the possibility
that TWA22\,AB is a member of the BPMG. With an M$6\pm1$ spectral
type, TWA22\,AB is probably close to the Li-depletion boundary (LDB) of
TWA or $\beta$ Pic, which could possibly explain a significantly stronger
EW(Li) than those observed for early-M dwarfs of these two
associations. We can also notice that the $\lambda$ 6708 \AA~line shows some variations between the S03 and M08 measurments. In addition, the observed EW(H$_\alpha$) of TWA22, used as a second indicator of youth, is compatible with those of GJ799 A
and B, M4.5 members of $\beta$ Pic \citep{2006ApJ...648.1206J}.

Finally, the projected position of TWA22\,AB reveals that the system is
isolated from other members of TWA. Its distance is more compatible with the mean distance of the BPMG members. \cite{2009Teixeira} have
recently measured the proper motion, the trigonometric parallax and the mean radial velocity of
TWA22\,AB. They determined for the first time the heliocentric space motion of TWA22\,AB.  From a detailed kinematic analysis they did not ruled out TWA22 from TW Hydrae but they demonstrated that it was a more probable member of the BPMG.

%\begin{table}[t]
%\caption{TWA22\,AB heliocentric space motion compared with those of
%TWA and $\beta$ Pic } % title of Table
%\label{tab:table6}
%\centering
%\begin{tabular*}{\columnwidth}{@{\excs}lllll}     % 7 columns
%\hline \hline        % inserts double horizontal lines
%Name        & U              &   V           &      W         \\
%            & (km.s$^{-1}$)  & (km.s$^{-1}$)& (km.s$^{-1}$)   \\  
%\hline  \noalign{\smallskip}         
%TWA22\,AB   & $-7.9\pm0.2$   & $-17.5\pm0.4$ & $-8.9\pm0.1$   \\
%\noalign{\smallskip}
%TWA         & $-10.5\pm0.9$   & $-18.0\pm1.5$ & $-4.9\pm0.9$   \\
%$\beta$ Pic & $-10.0\pm2.0$   & $-15.8\pm0.8$ & $-9.1\pm1.0$   \\
%\hline                                   %inserts single line
%\end{tabular*}
%\end{table}

%                                     Two column figure (place early!)
%______________________________________________ Gamma_1 (lg rho, lg e)
   
\section{Conclusions}

NACO resolved for the first time the young object TWA
22 as a tight binary with a projected separation of 1.76 AU. 80 \% of
the binary orbit was covered during a 4 years observation program
conducted with this instrument. We inferred a 220 $\pm$ 21 $M_{Jup}$
total mass for the system and we obtained the individual magnitudes of
each component in the near infrared. This places TWA22 A and B at the
substellar boundary. We complete the characterization of the system
components with medium resolution individual SINFONI spectra in the
\textit{J}, \textit{H} and \textit{K} bands. Our spectra were compared
with empirical library of young and field M dwarfs. We derived a M6
$\pm$ 1 spectral type from continuum fitting, spectral indexes and
equivalent widths. Spectral templates were also used to estimate
$T_{eff}$=2900 $\pm$ 200 K for TWA22 A and $T_{eff}$=2900
$^{+200}_{-100} $ K for TWA22 B while the surface gravity was
constrained to 4.0 $<$ \textit{log(g)} $<$ 5.5 dex. These fundamental properties can be directly compared with commonly used
evolutionary tracks provided that the age of the system is known accurately.

The age of TWA22 was still a mater of debate at the beginning of our study. TWA22 was reported as member of the young association TW Hydrae, and alternatively as a possible member of the BPMG.  At the age of TW Hydrae and BPMG, the new and precious benchmark brought by this system seems to point an underestimation of the predicted mass from our photometry.  However, the dynamical mass appears correctly estimated by the models if we consider a 30 Myr old system. This led us reconsider the membership of TWA22. 

While the spectroscopy tends to confirm the youth of this system, a recent kinematic study rejected TWA22 as a member of the TW Hydrae and of the 30 Myr old Tucana-Horologium associations. It didn't exclude the membership of TWA22 to the BPMG. Also, we can not rule out the possibility that TWA22 is not associated with any of these moving groups. \\

Finally, we does not exclude firmly that TWA22\,AB component could be in fact unresolved binaries with coplanar inner orbits characterized by semi-major axis lower than 0.4 AU. The models predictions would match the measured dynamical mass of a triple or quadruple system. In this context, future monitoring of TWA22\,AB with improved angular resolution could allow the resolution of the hypothetical inner binaries.

%Efforts to constrain the age of this young system need to be pursued
%in a near future. The next challenge could be to obtain visible
%spectra of individual components to confirm the infrared spectral
%types and to better constrain the surface gravities from an exended
%knowledge of the SED.  Rotational velocities could possibly test the
%membership of this singular system \citep{2005MNRAS.357.1399L}.

\begin{acknowledgements}
We thank the referee for an excellent and thorough review, which helped to improve our manuscript greatly. We thank the ESO Paranal staff for performing the service mode observations. We also acknowledge partial financial support from the \textit{Agence National de la Recherche} and the \textit{Programmes Nationaux de Planétologie et de Physique Stellaire} (PNP \& PNPS), in France. We are grateful to Andreas Seifahrt, Laird Close and Eric Nielsen, Catherine L. Slesnick, Nadya Gorlova, Katelyne N. Allers  and Nicolas Lodieu for providing their spectra. This work would have not been possible without the NIRSPEC and UKIRT libraries provided by Ian S. McLean, Michael C. Cushing and John T. Rayner. We also would like to thank Peter H. Hauschildt, France Allard and Isabelle Baraffe for their inputs on evolutionary models and synthetic spectral libraries. Finally, we thank Carlos Torres, Michael Sterzik and Ben Zuckerman, who gave use precious insights for the discussion. 
\end{acknowledgements}

\bibliographystyle{aa}
\bibliography{bibliography}

%\begin{thebibliography}{}
%
%  \bibitem[1966]{baker} Baker, N. 1966,
%      in Stellar Evolution,
%      ed.\ R. F. Stein,\& A. G. W. Cameron
%      (Plenum, New York) 333
%
%   \bibitem[1988]{balluch} Balluch, M. 1988,
%      A\&A, 200, 58
%
%   \bibitem[1980]{cox} Cox, J. P. 1980,
%      Theory of Stellar Pulsation
%      (Princeton University Press, Princeton) 165
%
%   \bibitem[1969]{cox69} Cox, A. N.,\& Stewart, J. N. 1969,
%      Academia Nauk, Scientific Information 15, 1
%
%   \bibitem[1980]{mizuno} Mizuno H. 1980,
%      Prog. Theor. Phys., 64, 544
%   
%   \bibitem[1987]{tscharnuter} Tscharnuter W. M. 1987,
%      A\&A, 188, 55
%  
%   \bibitem[1992]{terlevich} Terlevich, R. 1992, in ASP Conf. Ser. 31, 
%      Relationships between Active Galactic Nuclei and Starburst Galaxies, 
%      ed. A. V. Filippenko, 13
%
%   \bibitem[1980a]{yorke80a} Yorke, H. W. 1980a,
%      A\&A, 86, 286
%
%   \bibitem[1997]{zheng} Zheng, W., Davidsen, A. F., Tytler, D. \& Kriss, G. A.
%      1997, preprint
%\end{thebibliography}

\end{document}